\documentclass{article}

\usepackage{arxiv}

\usepackage[utf8]{inputenc}  
\usepackage[T1]{fontenc}     
\usepackage{hyperref}        
\usepackage{url}             
\usepackage{booktabs}        
\usepackage{amsmath}
\usepackage{amsfonts}        
\usepackage{nicefrac}        
\usepackage{microtype}       
\usepackage{cleveref}        
\usepackage{lipsum}          
\usepackage{graphicx}
\usepackage{natbib}
\usepackage{doi}

\title{An efficient and low-divergence method for generating inhomogeneous and anisotropic turbulence with arbitrary spectra}

\renewcommand{\headeright}{Research Paper}
\renewcommand{\undertitle}{Research Paper}

\renewcommand{\shorttitle}{%
  An Efficient \& Low-Divergence Method for Generating Turbulence}

\author
{%
  \href{https://orcid.org/0000-0001-6232-690}{\includegraphics[scale=0.06]{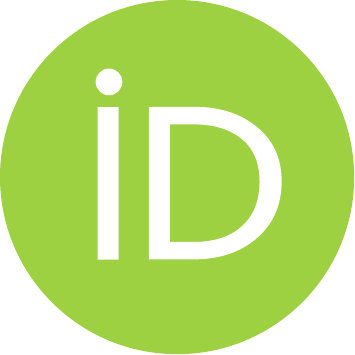}\hspace{1mm}Hao~Guo} \\
	Institute of Engineering Thermophysics \\
	Tsinghua University \\
	Beijing, China \\
	\texttt{guoh19@mails.tsinghua.edu.cn} \\
  \AND
  Peixue~Jiang \\
  Institute of Engineering Thermophysics \\
  Tsinghua University \\
	Beijing, China \\
  \texttt{jiangpx@tsinghua.edu.cn} \\
	\And
  Yinhai~Zhu\thanks{Corresponding author.} \\
	Institute of Engineering Thermophysics\\
	Tsinghua University \\
	Beijing, China \\
	\texttt{yinhai.zhu@tsinghua.edu.cn} \\
	\And
	Lin~Ye \\
  Institute of Engineering Thermophysics\\
	Tsinghua University \\
	Beijing, China \\
	\texttt{yelin@mail.tsinghua.edu.cn} \\
}

\graphicspath{ {figures/} {./} }

\hypersetup
{%
  colorlinks = true,
  urlcolor   = blue,
  citecolor  = black,
  pdftitle={An efficient and low-divergence method for generating inhomogeneous and anisotropic turbulence with arbitrary spectra},
  pdfsubject={Fluid dynamics},
  pdfauthor={Hao~Guo, Peixue~Jiang, Yinhai~Zhu},
  pdfkeywords={Spectrum-based method, Divergence-free, Inhomogeneous and anisotropic turbulence},
}

\begin{document}

\maketitle

\begin{abstract}
	In this article, we propose a divergence-free method for the generation of inhomogeneous and anisotropic turbulence. Based on the idea of correlation reconstruction, the method uses the Cholesky decomposition matrix to re-establish the turbulence correlation functions, which avoids the time-consuming procedure that solves eigenvalues and eigenvectors in every location needed by the coordinate transformation in the conventional method and thus reduces the computational complexity and improves the efficiency of generating synthetic turbulence. Through adjusting the generation strategy of specific random vectors, the proposed method, which is based on the classical spectrum-based method widely used to generate uniform isotropic turbulence, can obtain inhomogeneous and anisotropic turbulence with a relatively low divergence level in practice with almost no additional computational burden. There are two versions of this new method: the shifter version and the inverter version. The shifter-version method strictly guarantees the divergence-free result caused by both inhomogeneity and anisotropy. This version of the method also recovers to the classical spectrum-based method correctly when the anisotropy and inhomogeneity of the target turbulence are gradually reduced. The initial turbulence field generated using the shifter version method does not decay in subsequent calculations that solve the completed Navier--Stokes equations and can rapidly develop to be real turbulence that meets the statistical requirements. Further improving computation procedures, the inverter-version method solves the problems of anisotropy degeneration and turbulence kinetic energy reduction that may occur in the direct mapping operation of spherical distribution vectors and strictly corrects the divergence-free error caused by anisotropy. After thorough testing and analysis, we found that the increase in the practical level of divergence induced by anisotropy is much more significant than that induced by inhomogeneity in many common turbulent flows. When dealing with engineering problems, ignoring the divergence-free error caused separately by the inhomogeneity will hardly have an apparent effect on the actual divergence level of the synthetic turbulence. Therefore, the inverter version method ignores the inhomogeneity error term and only strictly corrects the error term due to anisotropy, further reducing the computational cost. Both versions of the method are highly efficient, easy to implement, and compatible with high-performance computing. Suitable for providing high-quality initial or boundary conditions for scale-resolving turbulence simulations with large grid numbers (such as direct numerical simulation or large eddy simulation), this method can be quickly implemented either based on various open-source CFD codes or common commercial CFD software.
\end{abstract}

\keywords{Spectrum-based method \and Divergence-free \and Inhomogeneous and anisotropic turbulence}

\section{Introduction}

With the rapid development of high-performance computing (HPC) and related hardware capabilities, scale-resolving turbulence simulations, such as direct numerical simulation (DNS) and large eddy simulation (LES), have been gradually applied to increasingly complex geometries and working conditions in recent years \citep{LESbook}. Although great progress has been made, acquiring initial or boundary conditions of general turbulence remains an unavoidable problem and has not been perfectly solved for decades \citep{Tabor:2010,Dhamankar:2017}.

The acquisition of initial conditions and boundary conditions for highly-resolved turbulence simulation can be roughly divided into two categories, i.e. the database method and the synthetic turbulence method. Detailed descriptions can be found in the review literature of \citet{Dhamankar:2017}. By directly using experimental data, storing results of DNS, or constructing real-time mapping from additional computational domains \citep{Lund:1998,EA:2003}, the database method can obtain field data that are consistent with or close to the real turbulence, with high precision and accuracy. The main drawback of this type of method is that the scope of the application is very limited. To preserve the original information of real turbulence as much as possible, the data storage and access burden of these methods are usually very large, and sometimes the corresponding computational cost is even of the same order as the main computational domain of focussed turbulent flows. Most methods are not well adapted to parallel programming. Moreover, a single method is often only applicable to a specific type of turbulence, which is difficult to meet the requirement of generating more complex turbulence in engineering. Although some scholars have tried to build turbulence generators based on databases through deep learning \citep{Yousif:2022}, the synthetic turbulence method is undoubtedly a more direct and effective method to solve this general problem. The idea of the synthetic turbulence method is to synthesize turbulence based on certain information to provide necessary initial conditions and boundary conditions. It covers a wide range of specific method types, including spectral-representation-based method \citep{Kraichnan:1970,Smirnov:2001}, digital-filter-based method \citep{Klein:2003,Kempf:2005,Kim:2013}, synthetic-eddy method (SEM) \citep{Benhamadouche:2006,Jarrin:2006,Mathey:2006,Subbareddy:2006} and diffusion-based method \citep{Kempf:2005}. Due to its valid spatial correlation and scale distribution property that can realize arbitrary spectra, the spectrum-based method is widely applied to many in-house codes and general computational-fluid-dynamic (CFD) softwares. For example, both the open-source code OpenFOAM \citep{FOAM} and the commercial software \citet{Fluent} have implemented some versions of the spectrum-based method.

One of the most important requirements for synthetic turbulence is to ensure the divergence-free or solenoidal property condition in an incompressible flow. On one hand, synthetic incompressible turbulence that does not fulfill this condition may rapidly decay or cause an excessive pressure fluctuation \citep{Poletto:2013}. On the other hand, for compressible flows, the artificially introduced velocity divergence will produce spurious noise, which may interfere with other shocks or expansion waves in high-speed flow \citep{Dhamankar:2017}. Many types of synthetic turbulence methods, such as digital filters \citep{Klein:2003} and synthetic eddy methods \citep{Jarrin:2006}, in their widely used classical form, do not naturally generate divergence-free results because of the focus mainly on producing spatial/temporal correlations or physical statistics precisely. Although some studies have tried to improve the digital filter method or the synthetic eddy method to meet the solenoidal requirements \citep{Kim:2013,Poletto:2013}, these improvements usually come with a price due to the limitations of the methodological framework. Most improved versions either lose some of the advantages that the classical version preserves or attach more limitations to the method for ensuring the divergence-free property. For example, the improved method of Kim et al. cannot meet the requirements of keeping a constant mass flux and divergence-free at the same time \citep{Kim:2013}. The method proposed by \citet{Kempf:2005}, if the projection step \citep{Lee:1992} is employed to correct the divergence error, not only become problematic dealing with boundary conditions but also results in the deviation of desired correlations under some circumstances. The vortex method (VM) using the Biot-Savart law can guarantee divergence-free conditions only if there are no streamwise fluctuations \citep{Mathey:2006}.

Different from other synthetic method types, the classical form of the spectrum-based method generating homogeneous and isotropic turbulence strictly produces a divergence-free result. However, subsequent improvements extending it to cover inhomogeneous and anisotropic turbulence have lost this solenoidal property more or less. The method proposed by \citet{Smirnov:2001} is one of the most widely used spectrum-based methods that handle inhomogeneity and anisotropy. This method uses a coordinate transformation and scaling operation to achieve this inhomogeneity-and-anisotropy extension, but the divergence of the resulted field cannot ensure the divergence-free condition anymore if the target turbulence is not homogenous or isotropic. A recent study has shown that even an approximate zero divergence claimed by Smirnov et al. does not hold in many cases \citep{Yu:2014}. Moreover, to obtain the orthogonal matrix for coordinate transformation, all eigenvalues and eigenvectors of the Reynolds stress tensor need to be calculated for each grid point, which significantly increases the computational cost of this type of method. Based on the extension version of Smirnov et al., Yu et al., also introducing coordinate transformation, improve the method to ensure divergence-free results by using curl of a vector potential field \citep{Yu:2014}. Although their method fulfills the solenoidal requirement, the time-consuming eigenvalue and eigenvector calculation is also inherited from Smirnov's method since coordinate transform matrixes are still needed for every gird point. Besides, the introduction of the idea of vector potential makes their method change fundamentally compared with the conventional spectrum-based method and the computational complexity is significantly increased, which means that most of the existing spectrum-based method code frameworks can hardly be modified to implement this new version and therefore greatly limiting the practical application of it.

In this paper, a new divergence-free spectrum-based method is proposed, which is based on Cholesky decomposition for correlation reconstruction instead of coordinate transform to solve the problem of inhomogeneity and anisotropy extension. Its basic idea for acquiring divergence-free turbulence is to correct the calculation strategy of specific defective steps according to divergence-free error terms resulting from correlation reconstruction, so it completely inherits the framework and advantages of the classical spectrum-based method. The algorithm of this new method is quite concise, requiring only two steps of the original one modified to ensure the divergence-free property in the inhomogeneous and anisotropic flow. Since eliminating conventional coordinate transformation, no eigenvalue and eigenvector solving algorithms are involved, which achieves a faster computation speed when dealing with a large number of grid points, especially in high-resolution DNS/LES. The effectiveness of the proposed method has been verified by four types of test cases: homogeneous and isotropic turbulence, anisotropic turbulence, inhomogeneous turbulence, and typical inhomogeneous and anisotropic turbulence of channel flow.

The structure of this paper is organized as follows: Section \ref{sec:method} explains the theory and derivation of the proposed method in detail. Section \ref{sec:algorithm} analyzes the algorithms and computational complexity of different versions. In Section \ref{sec:results}, representative cases of four types are employed to verify the method and test its practical performance thoroughly.

\section{Theory and methodology}
\label{sec:method}

\subsection{Methods for generating homogeneous and isotropic turbulence}
\label{ssec:origin}

Adopting \citet{Bechara:1994}'s notation, the velocity vector in Cartesian space for a single sampling can be expressed in Fourier series as
\begin{equation}
\label{eq:main:origin}
  u_i\left(x_{(k)},t\right) =
    \sum_n 2\, p^{(n)}
    \cos\left(
      \kappa^{(n)}_j x_j + \omega^{(n)} t + \varphi^{(n)}
    \right)
    \sigma^{(n)}_i,
\end{equation}
where, the superscript $^{(n)}$ represents the nth Fourier mode and does not participate in the tensor summation. Mode phase $\varphi^{(n)} \sim \mathrm{U}(0,2\pi)$ and mode amplitude $p^{(n)} = \sqrt{E_{k}(\kappa^{(n)}) \Delta \kappa^{(n)}}$. The energy spectrum can be obtained using experimental data, such as CBC data \citep{CBC:1971}, which are widely used in calculations of homogeneous and isotropic turbulence decaying. In the algorithm for generating the fluctuation velocity, when computing each mode, the unit wave vector $\hat{\kappa}_i^{(n)}$ of the mode is first obtained by generating a random unit vector. Symbols $\hat{}$  represent the corresponding unit vector after normalization, e.g. $\kappa_i^{(n)} = \kappa^{(n)} \hat{\kappa}_i^{(n)}$. Vector product or cross product is then used to obtain the mode direction that is perpendicular to the wave vector:
\begin{align}
\label{eq:perp:origin}
  \sigma^{(n)}_i &=
  \left(
    \varepsilon_{ijk} \xi^{(n)}_j \hat{\kappa}^{(n)}_k
  \right)_\mathrm{normalize},
\end{align}
where, $\xi^{(n)}$ is a random unit vector following the spherical distribution as well. If let
$\alpha^{(n)}\left(x_{(k)},t\right) =
  \kappa^{(n)}_j x_j + \omega^{(n)} t + \varphi^{(n)}$,
we can get $\alpha_{,j}^{(n)} = \kappa_i^{(n)}$ and $\partial_t \alpha^{(n)} = \omega^{(n)}$.

For a given time instant, the divergence of the turbulent velocity field can be expressed as
\begin{align}
  u_{i,\,i} &=
    \sum_n 2\, p^{(n)} \sin \alpha^{(n)} \kappa^{(n)}_i \sigma^{(n)}_i.
\end{align}

As long as the wave vector and direction vecto	r are strictly perpendicular, it is ensured that the generated turbulence satisfies the divergence-free condition.

\subsection{Extented method to handle inhomogeneous and anisotropic turbulence}
\label{ssec:extent}

The generation method represented by Eq.\eqref{eq:main:origin} only accounts for homogenous and isotropic turbulence, and it needs some extension technique to cover inhomogeneous turbulence. A typical one is to introduce coordinate transformation and scaling operations \citep{Smirnov:2001,Yu:2014}. However, the orthogonal matrix computation for coordinate transformation requires the calculation of all eigenvalues and eigenvectors of the Reynolds stress tensor at each spatial point where synthetic turbulence is needed. As a consequence, the computational cost has increased significantly. To ensure the efficiency of computation, this paper adopts the correlation reconstruction transformation based on Cholesky decomposition matrixes. i.e., the velocity is generated by the following formula:
\begin{align}
  \label{eq:ext:v}
  v_i\left(x_{(k)},t\right) &=
    \sum_n 2\, q^{(n)}
    \cos\left(
      \kappa^{(n)}_j x_j + \omega^{(n)} t + \varphi^{(n)}
    \right)
    \sigma^{(n)}_i,  \\
  \label{eq:ext:u}
  u_i\left(x_{(k)}, t\right) &=
    L_{ij}\left(x_{(k)}, t\right) v_j,
\end{align}
where $q^{(n)} = p^{(n)}/u_t$. $u_t$ is the characteristic velocity of turbulence corresponding to the integration scale and can be calculated through $\sqrt{ \frac13{\langle u_i u_i \rangle} }$.

The correlation reconstruction matrix $L_{ij}$ is obtained based on the Cholesky decomposition of the Reynolds stress tensor $R_{ij}$, i.e. $R_{ij} = L_{ik} L_{jk}$. Therefore,
\begin{equation}
\label{eq:Cholesky}
  \begin{gathered}
      L_{11} = \sqrt{R_{11}},\;
      L_{21} = \frac{R_{21}}{L_{11}},\;
      L_{22} = \sqrt{R_{22} - L_{21}^2},  \\
      L_{31} = \frac{R_{31}}{L_{11}},\;
      L_{32} = \frac{R_{32} - L_{31} L_{21}}{L_{22}},\;
      L_{33} = \sqrt{R_{33} - L_{31}^2 - L_{32}^2},  \\
      L_{12} = L_{13} = L_{23} = 0.
  \end{gathered}
\end{equation}

\subsection{Divergence-free errors caused by correlation reconstruction}

Although the correlation reconstruction matrix is relatively concise, a deficiency of this extended method is that the divergence is no longer strictly 0. After derivation, the velocity divergence at this point is
\begin{equation}
\label{eq:errors}
  \begin{split}
    u^{(n)}_{i,\,i} =
      2 \sum_n
      \Bigg[
      \underbrace
      {
          \cos\alpha^{(n)} q^{(n)}_{,\,i} L_{ij} \sigma^{(n)}_j
      }_{Er_1}
      &+
      \underbrace
      {
          \cos\alpha^{(n)} q^{(n)} L_{ij} \sigma^{(n)}_{j,\,i}
      }_{Er_2}
      -
      \underbrace
      {
          \sin\alpha^{(n)} q^{(n)} L_{ij} \sigma^{(n)}_{j} \kappa^{(n)}_{k,\,i} x_{k}
      }_{Er_3} \\
      &+
      \underbrace
      {
          \cos\alpha^{(n)} q^{(n)} L_{ij,\,i} \sigma^{(n)}_j
      }_{Er_4}
      -
      \underbrace
      {
          \sin\alpha^{(n)} q^{(n)} \kappa_i L_{ij} \sigma^{(n)}_{j}
      }_{Er_5}
      \Bigg]
  \end{split}
\end{equation}
There are five error terms in Eq.\eqref{eq:errors} responsible for the non-zero divergence in inhomogeneous and anisotropic turbulence. The first term $Er_1$ arises from the inhomogeneity of the amplitude $q^{(n)}$ and is present in almost all the improved spectrum-based methods for inhomogeneous turbulence. Since it is calculated from the normalized energy spectrum, $Er_1$ is mainly affected by the change of turbulence integral scale in space. However, the effect of this term is very small compared with others and can be ignored in practice. The second and third terms are strictly zero in the previous method. After introducing modifications, only if the unit direction vector $\sigma_i^{(n)}$ or wave vector $\kappa_i^{(n)}$ exhibits strong variation will introduce corresponding errors. Since $Er_2$ and $Er_3$ are essentially bound to additional adjustment, it is extremely difficult to design the correction in advance so that these two terms become zero naturally. The good news is that most of the improvements cause both of these errors to be so small that they are negligible in practical calculations.

The increase in divergence magnitude produced by the correlation reconstruction operation is mainly contributed by terms $Er_4$ and $Er_5$, so the correction method to strictly ensure the divergence-free property of result turbulence starts with these two error terms. The fourth term $Er_4$ represents the inhomogeneity of the correlation reconstruction matrix, which is essentially caused by the spatial distribution of the Reynolds stress tensor field. When the Reynolds stress $R_{ij}$ is uniform in the computational domain, $Er_4 = 0$. The fifth term $Er_5$ is the anisotropy influence of the correlation reconstruction, which is essentially caused by the anisotropy of the local $R_{ij}$. When $R_{ij}$ is isotropic, the fifth term degenerates to the form $\sin\alpha^{(n)} q^{(n)} \tfrac{1}{3} L_{kk} \kappa_i^{(n)} \delta_{ij} \sigma_i^{(n)} = 0$, and $Er_5$ vanishes.

\subsection{Divergence-free correction: shifter version}

Note that the direction vector $\sigma_i^{(n)}$ exists in the same form in $Er_4$ and $Er_5$, so that when satisfies
\begin{equation}
  \left(
    \cos\alpha^{(n)} L_{ij,i} - \sin\alpha^{(n)} \kappa_i^{(n)} L_{ij}
  \right)
  \sigma_j^{(n)} =
    \tilde{\kappa}_j^{(n)} \sigma_j^{(n)} = 0,
\end{equation}
i.e., $\sigma_j^{(n)}$ and $\kappa_i^{(n)}$ are perpendicular to each other, the divergence of the generated turbulence is strictly 0. Therefore, by improving the construction strategy of the direction vector, we obtain the shifter version of the divergence-free spectrum-based method suitable for the generation of inhomogeneous and anisotropic turbulence. Firstly, the intermediate wave vector is calculated by the following formula:
\begin{equation}
\label{eq:shifter}
  \tilde{\kappa}_j^{(n)} =
    \cos\alpha^{(n)} L_{ij,i} - \sin\alpha^{(n)} \kappa_i^{(n)} L_{ij}.
\end{equation}
The mode direction vector perpendicular to the intermediate wave vector is then constructed. The shifter version of this divergence-free method simultaneously corrects for the correlation reconstruction errors caused by inhomogeneity and anisotropy and guarantees fully solenoidal characteristics of generated turbulence. However, in practical application, we noticed that a few issues need attention as well in the shifter version, acquiring further improvement accordingly.

1) There are trigonometric functions in Eq.\eqref{eq:shifter}, and the intermediate wave vector shifts in the plane by the divergence of the reconstruction matrix $L_{ij,i}$ and the vector $\kappa_i^{(n)} L_{ij}$ with different spatial positions, which is why this version is called “shifter”. Under the circumstance that the inhomogeneity of the turbulent field is very small ($L_{ij,i}$ is close to or strictly equal to 0), the existence of $\sin\alpha^{(n)}$ function will lead to frequent flipping of the supposed constant vector $\sigma_i^{(n)}$ if the perpendicular vector is still constructed by the procedure of Eq.\eqref{eq:perp:origin}. This drawback not only leads to a serious deviation of the synthetic turbulence from the given energy spectrum but also means that the method is unable to recover to the original version presented in Section \ref{ssec:origin} correctly when dealing with homogeneous turbulence. The reason for this flipping issue is that the restriction brought by theory only gives a perpendicular line but the direction vector still has two options. The final direction depends on the algorithm employing vector product operation numerically. However, the one-time vector product directly introduces the positive and negative effects of trigonometric functions, resulting in the direction vector jumping between the two possibilities according to spatial positions. To overcome this drawback, we introduce two times vector product operation in the direction computation step, i.e., substituting Eq.\eqref{eq:perp:origin} with the following procedure:
\begin{align}
  \label{eq:perp:1}
  \zeta^{(n)}_i &=
  \left(
    \varepsilon_{ijk} \xi^{(n)}_j \tilde{\kappa}^{(n)}_k
  \right)_\mathrm{normalize},  \\
  \label{eq:perp:2}
  \sigma^{(n)}_i &=
  \left(
    \varepsilon_{ijk} \tilde{\kappa}^{(n)}_j \zeta_j^{(n)}
  \right)_\mathrm{normalize}.
\end{align}
This treatment consists of two vector products, essentially calculating the projection of the random unit vector $\xi_i^{(n)}$ on the plane perpendicular to $\kappa_i^{n}$. Moreover, this treatment is more reasonable in physics than the one-time vector product version because the quantity containing the permutation symbol $\varepsilon_{ijk}$ is not a tensor in essence (not satisfying Galilean invariance in a mirror coordinate transformation).

2) Although the correction method of the shifter version strictly ensures the solenoidal condition in a larger application, it changes the generation process of the direction vector and introduces additional inhomogeneity and anisotropy. As a result, the covariance matrix of intermediate velocity $v_i$ will deviate from the identity matrix leading to Reynolds stress departure from the desired one. This deviation will not only exhibits a degenerated anisotropy of the given correlation $R_{ij}$ but also reduce turbulent kinetic energy (TKE) magnitude. The TKE reduction can be almost eliminated through scaling according to two types of anisotropy intensity. But for the anisotropy degeneration issue, the solution is not that simple: a special construction strategy that eliminates the change due to anisotropy from the mode direction vector completely should be employed, giving rise to an inverter version divergence-free method presented in Section \ref{ssec:inverter}. The fundamental reason for this deviation is that the random wave vector originally following a spherical distribution cannot maintain this characteristic after a series of specific transformations. Moreover, we noticed that this deviation can hardly be controlled once produced and depends on the way how the random vector is generated (e.g. there exists a normalization step or not). Further discussion concerning anisotropic degeneration and TKE reduction issues are presented in detail in Section \ref{ssec:rHI}.

In summary, the shifter version of the correction method strictly ensures the divergence-free property of inhomogeneous and anisotropic turbulence. Although the generated turbulence field will have a deviation in the correlation function, we found that the shifter version, preserving a solid spatial correlation, is qualified for the generation of turbulent boundary and initial field but requires an increased recovery length (boundary condition) or recovery time (initial condition), which is still smaller than that of some other methods where a decay in turbulence is observed. Due to the effectiveness of spatial correlation, the initial deviation of synthetic turbulence produced by the shifter version method can quickly develop and recover to the correct state without decaying.

\subsection{Divergence-free correction: inverter version}
\label{ssec:inverter}

Note that in Eq.\eqref{eq:errors}, with a large wave vector $\kappa_i^{(n)}$ or a smaller length scale, the reconstruction error term $Er_5$ of a specific Fourier mode caused by anisotropy tends to have a large magnitude as well while the inhomogeneity error term $Er_4$ remains unaffected. It is expected that when the wave number $\kappa^{(n)} = \|\kappa_i^{(n)}\|$ is large enough to exceed a critical value, the overall divergence-free error caused by the correlation reconstruction will be dominated by $Er_5$. On this occasion, only accounting for $Er_5$ can also lead to a significantly low divergence level, which is one of the basic ideas behind the inverter version correction.

Another idea driving to inverter version is to overcome degenerated anisotropy and TKE reduction explained previously. Considering the possible correlation deviation caused by modifying direction vectors $\sigma_i^{(n)}$, in the inverter version, we abandon the conventional idea of previous research \citep{Saad:2016} and the shifter version to obtain $\sigma_i^{(n)}$ by first calculating inhomogeneous temporary wave vector field according to the uniform wave vector. Instead, the uniform $\sigma_i^{(n)}$ of each mode is obtained first, and the temporary direction vector is calculated by the following formula:
\begin{equation}
\label{eq:inverter}
  \tilde{\sigma}_i^{(n)} = L_{ij} \sigma_j^{(n)}.
\end{equation}
The final wave vector direction $\hat{\kappa}_i^{(n)}$ perpendicular to $\tilde{\sigma}_i^{(n)}$ is then obtained. Since there is no trigonometric function in Eq.\eqref{eq:inverter}, either using the vector product once or twice to calculate the perpendicular vector works. The inverter version of the divergence-free spectrum-based method completely avoids the correlation deviation reflected in degenerated anisotropy and damped TKE, and transfers all possible effects to the wave vector of each mode, which has exhibited a negligible effect on the energy spectrum during our practical computation test.

For homogeneous anisotropic turbulence, this version of the generation method strictly corrects the divergence-free error; for inhomogeneous anisotropy, the inverter version method can correct $Er_5$ effectively as well. Only in the synthesis of inhomogeneous and isotropic turbulence, does the inverter version return to the primitive extended method naturally. But such types of turbulent flows are very rare in engineering applications. Apparently, the correction performance of the inverter version depends on the effect of ignoring $Er_4$. Investigating Eq.\eqref{eq:errors}, it can be found that when
\begin{equation}
\label{eq:sig:ll1}
  \frac{
    \|L_{ij,i}\|_{\mathrm{2-norm}}
  }{
    \kappa^{(n)} \| L_{ij} \|_{\mathrm{2-norm}}
  }
    \ll 1,
\end{equation}
$Er_4$ is negligible compared to $Er_5$. The physical meaning behind Eq.\eqref{eq:sig:ll1} is that the spectrum-based method essentially superimposes Fourier modes/series to construct/reproduce the turbulent fluctuations, but the largest scale that can be identified by the discrete Fourier transform is the largest spatial scale corresponding to the computational domain. The zero-wave-number mode is a constant without spatial distribution. Therefore, spatial inhomogeneity is the signal with wave numbers beyond the recognition range of the discrete Fourier series, belonging to $[0,2\pi/L_d]$. As the wave number increases, the micro-scale turbulence tends to be homogeneous and the influence of this small-wave-number signal becomes weaker. Therefore, it is the large-scale turbulence that is mainly affected by the macroscopic inhomogeneity, and the divergence-free errors due to the correlation reconstruction are also mainly concentrated in the large-scale modes close to the scale of the computational domain. Thus, there exists a critical mode $n_c$ where the effect of $Er_4$ can be ignored for arbitrary Fourier modes fulfilling $\kappa^{(n)} > \kappa^{(n_c)}$, i.e., the error term associated with spatial inhomogeneity generated by these modes can be regarded as zero. However, for realistic turbulence, its integral scale/energy-containing scale is usually in the large scale range as well. Whether $Er_4$ is negligible depends on the relative size of this critical mode and the integral scale of turbulence. If satisfying
\begin{equation}
\label{eq:sig:kk1}
  \frac{\kappa_c}{\kappa_l} =
    \frac{
      \|L_{ij,i}\|_{\mathrm{2-norm}}
    }{
      2\pi \sqrt{
        \rho \left(R_{ij}\right)
      }
    }
      \ll 1,
\end{equation}
the reconstruction divergence-free error due to the inhomogeneity is negligible. In Eq.\eqref{eq:sig:kk1}, $l$ is the integral scale of local turbulence and $\rho \left(R_{ij}\right)$ represents the spectral radius of local $R_{ij}$. As will be further explained and verified in Sections \ref{ssec:rI} and \ref{ssec:rAI}, the condition of Eq.\eqref{eq:sig:kk1} is usually satisfied in practical applications.

\subsection{Energy spectrum models}

\begin{figure}
	\centering
  \fbox{\includegraphics[width=8cm]{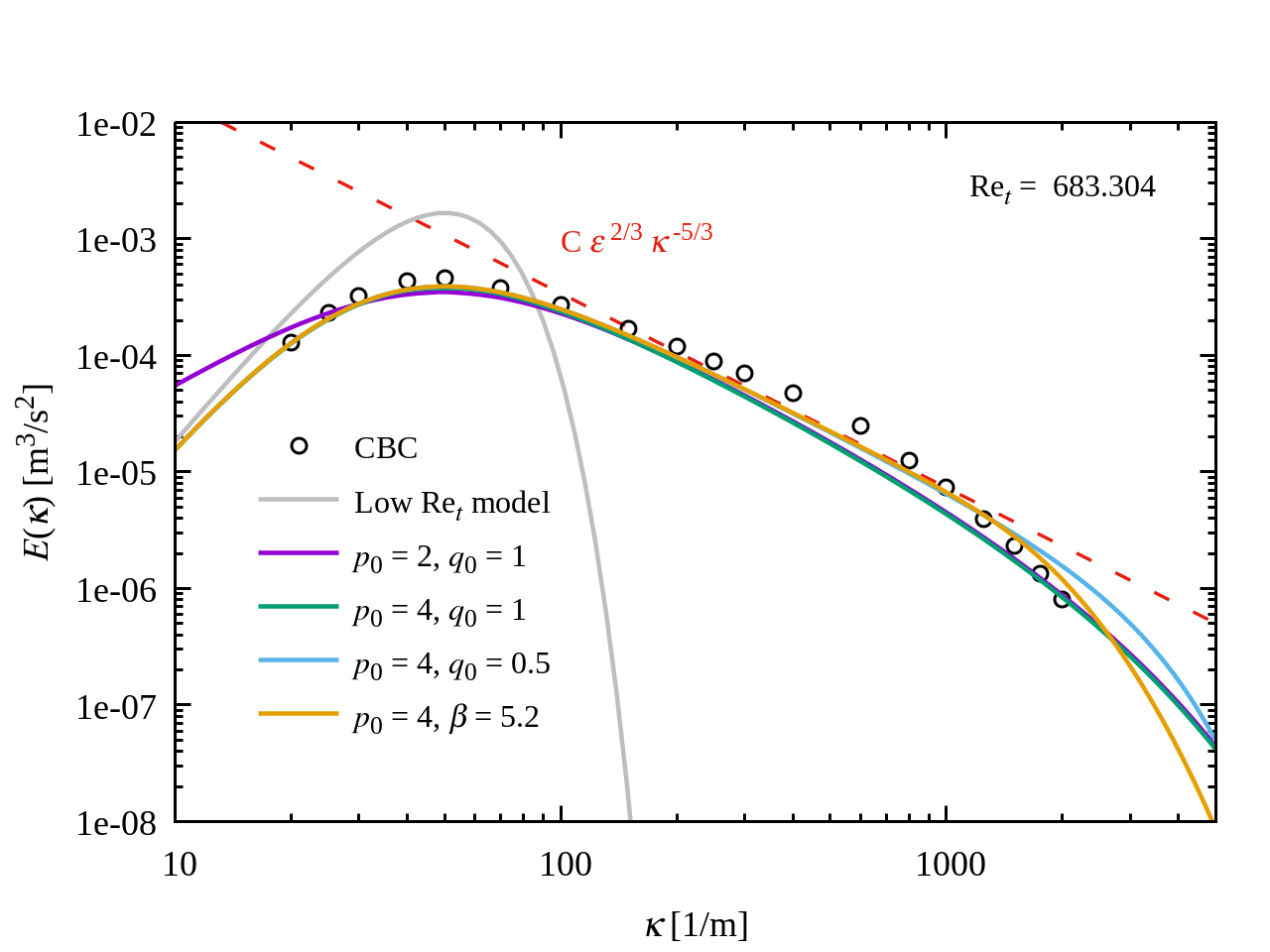}}
	\caption{Comparison of different energy spectrum models with experimental data.}
	\label{fig:spectra}
\end{figure}

One of the distinct advantages of the spectrum-based method is that it can produce turbulence based on arbitrary forms of energy spectra. In addition to some specific flows, which can be acquired by interpolating experimental data \citep{CBC:1971}, many existing studies employed various energy spectrum models to ensure the capability for complex flows \citep{Karman:1948,Pao:1965,Kraichnan:1970,Bailly:1999,Yu:2014}. Except for the low-Reynolds-number energy spectrum model used by \citet{Kraichnan:1970}, almost all high-Reynolds-number models can be expressed in a general form as follows:
\begin{equation}
\label{eq:spectra}
  E_k(\kappa) = C \varepsilon^{2/3} \kappa^{-5/3} f_l(\kappa l) f_\eta(\kappa \eta),
\end{equation}
where functions $f_l$ and $f_\eta$ describe the distribution form of the spectrum in energy-containing range and dissipation range respectively. We compared various energy spectrum models including the low-Reynolds-number model with the CBC spectrum data \citep{CBC:1971}, and the results are illustrated in Fig.\ref{fig:spectra}.

It can be found from Fig.\ref{fig:spectra} that the high-Reynolds-number model adopting the energy-containing range model of $p_0 = 4$ together with the dissipation range model of $\beta = 5.2$ exhibits the best performance, so this combination of the energy spectrum model is implemented in the code. Detailed information concerning certain parameters and the calculation of model coefficients can be found in Pope's book \citep{Pope:2001}.

\section{Algorithm}
\label{sec:algorithm}

In this section, specific implementation algorithms of the shifter and inverter version correction method are described respectively, and the corresponding computation complexity and runtime storage requirements are analyzed.

\subsection{The shifter version correction method}

A recipe for the shifter version is as follows:

\begin{enumerate}
  \item The correlation reconstruction tensor field $L_ij$ is calculated from Eq.\eqref{eq:Cholesky} based on the given Reynolds stress distribution.
  \item Generate random unit vectors $\hat{\kappa}_i^{(n)}$ and $\xi_i^{(n)}$ for each Fourier mode.
  \item Calculate the temporary or intermediate wave vector $\tilde{\kappa}_i^{(n)}$ from Eq.\eqref{eq:shifter}.
  \item The direction vector $\sigma_i^{(n)}$ is obtained based on $\xi_i^{(n)}$ by twice vector products using Eq.\eqref{eq:perp:1} and Eq.\eqref{eq:perp:2}.
  \item Calculate the intermediate velocity field $v_i$ through Eq.\eqref{eq:ext:v}.
  \item The final velocity field $u_i$ is calculated from Eq.\eqref{eq:ext:u} by the correlation reconstruction tensor $L_{ij}$.
\end{enumerate}

According to the above steps, compared with the earlier spectrum-based methods \citep{Kraichnan:1970,Bechara:1994} applicable to the generation of homogeneous and isotropic turbulence, the additional computational complexity mainly consists of the Cholesky decomposition of Reynolds stress field and the construction of intermediate wave vectors. Since the trigonometric functions in the latter will eventually be used in Eq.\eqref{eq:ext:v}, the construction of the intermediate wave vector in step 3 adds only a few multiplication operations. Although Cholesky decomposition needs to operate on the entire domain, it only needs to be performed once regardless of mode and does not involve any algorithm computing eigenvalues or eigenvectors. Compared with the algorithm requiring coordinate transformation and scaling operations, the efficiency of this proposed method is significantly improved. The above computational procedure is concise and easy to implement, only requiring a slight modification of the spectrum-based method code that generates homogeneous and isotropic turbulence to realize this new method.

The additional runtime memory usage in the shifter version of the proposed divergence-free method is only the storage of the correlation reconstruction tensor field $L_{ij}$ and is almost negligible. Moreover, since $L_{ij}$ is a lower triangular matrix, only 6 components need to be stored for each grid point in practice rather than 9 components of a typical orthogonal matrix. All the above steps use only local storage and local information, so the proposed method is easy to parallelize naturally and suitable for HPC of scale-resolving turbulence simulation.

\subsection{The inverter version correction method}

A recipe for the inverter version is as follows:

\begin{enumerate}
  \item The correlation reconstruction tensor field $L_ij$ is calculated from Eq.\eqref{eq:Cholesky} based on the given Reynolds stress distribution.
  \item Generate the unit direction vector $\sigma_i^{(n)}$ and random unit vector $\xi_i^{(n)}$ for each Fourier mode.
  \item Calculate the intermediate direction vector $\tilde{\sigma}_i^{(n)}$ from Eq.\eqref{eq:inverter}.
  \item The wave vector $\kappa_i^{(n)} = \kappa^{(n)} \hat{\kappa}_i^{(n)}$ is calculated by one-time or two-time cross-product operation using $\xi_i^{(n)}$.
  \item Calculate the intermediate velocity field $v_i$ through Eq.\eqref{eq:ext:v}.
  \item The final velocity field $u_i$ is calculated from Eq.\eqref{eq:ext:u} by the correlation reconstruction tensor $L_{ij}$.
\end{enumerate}

It can be seen that the difference between the inverter version and the shifter version is in steps 2--4, especially in step 3. The computation cost for the construction of the intermediate direction vector $\tilde{\sigma}_i^{(n)}$ using Eq.\eqref{eq:inverter} is further reduced compared to the construction of its intermediate wave vector $\tilde{\kappa}_i^{(n)}$ using Eq.\eqref{eq:shifter}. Similarly, in the inverter version of the correction method, the implementation of correlation reconstruction for inhomogeneity and anisotropy extension does not involve any coordinate transformation or eigenvalue computation algorithm, which significantly improves the computation speed, especially in HPC. Similar to the shifter one, the inverter version is concise and efficient and codes of some primitive spectrum-based methods require only a slight change to realize this divergence-free extension.

Same as the shifter version, the inverter version of the correction method only adds a small amount of memory usage to the original method, and these runtime storage are all field data that can be arranged in distributed machines with corresponding mesh. The algorithm only requires local information, so it is easy to perform parallelization in HPC,  suitable to provide both the initial condition and the boundary condition for high-cost LES or DNS.

\section{Results and discussion}
\label{sec:results}

In order to verify the effectiveness of the new method, a total of four types of verification calculations were carried out in this paper. The code implementation and corresponding CFD computations of each method are realized based on OpenFOAM \citep{FOAM}. The number of modes in all cases is 5000, and the interpolation method of logarithmically distributed modes is employed. Case 1 is a classical benchmark of homogeneous and isotropic turbulence decay, which is mainly used to illustrate the correct realization of the classical spectrum-based method and the effectiveness of the CFD codes, and to verify that the new method accurately reverts to the original method in the application of homogeneous and isotropic turbulence. Another important purpose of case 1 is to obtain the benchmark value of the divergence level at which the box-geometry-type case can be considered to achieve divergence-free in practical calculations under the same number of grids. Case 2 and case 3 similarly used the box geometry of case 1. Case 2 focuses on the performance of the method handling anisotropic turbulence, so homogeneous synthetic turbulence with different anisotropy types is examed in detail and the difference in performance producing desired statistics between the shifter version and the inverter version is compared. The anisotropy degeneration and turbulent kinetic energy decay issues produced by the shifter version are also studied in case 2. Case 3 concentrates on verifying the treatment of inhomogeneity. An inhomogeneous turbulence with distribution only in x direction and a wavelength of one computational domain size is assumed. The purpose of this artificial design is to keep the y-z plane for spatial averaging in a single realization/sampling in addition to ensemble averaging for multiple realizations/samplings. In case 3, mainly adopting the inverter version of the divergence-free method, the magnitude of reconstruction divergence-free errors caused by inhomogeneity and anisotropy are compared. Case 4 considers a typical application of inhomogeneous and anisotropic turbulence in engineering practice: channel flow. Firstly, it strictly analyzes whether the inhomogeneous divergence-free errors of the actual turbulence can be ignored. Case 4 is mainly carried out using the inverter version as well, comparing the turbulence generation using a uniform integral scale and a spatial-distributed integral scale, respectively, to investigate the effect of the inhomogeneity of the normalized energy spectrum. Finally, the proposed new method is applied to generate the initial field of the channel flow to verify the practical performance in a scale-resolving turbulence simulation.

In case 1 and case 4, large eddy simulations were carried out with second-order temporal and spatial discretizations. The sub-grid model of LES in case 1 was the WALE model \citep{Nicoud:1999}. In case 4, the sub-grid model was not employed because a mesh of DNS resolution was used.

\subsection{Homogeneous and isotropic turbulence}
\label{ssec:rHI}

\begin{figure}
	\centering
  \fbox{\includegraphics[width=8cm]{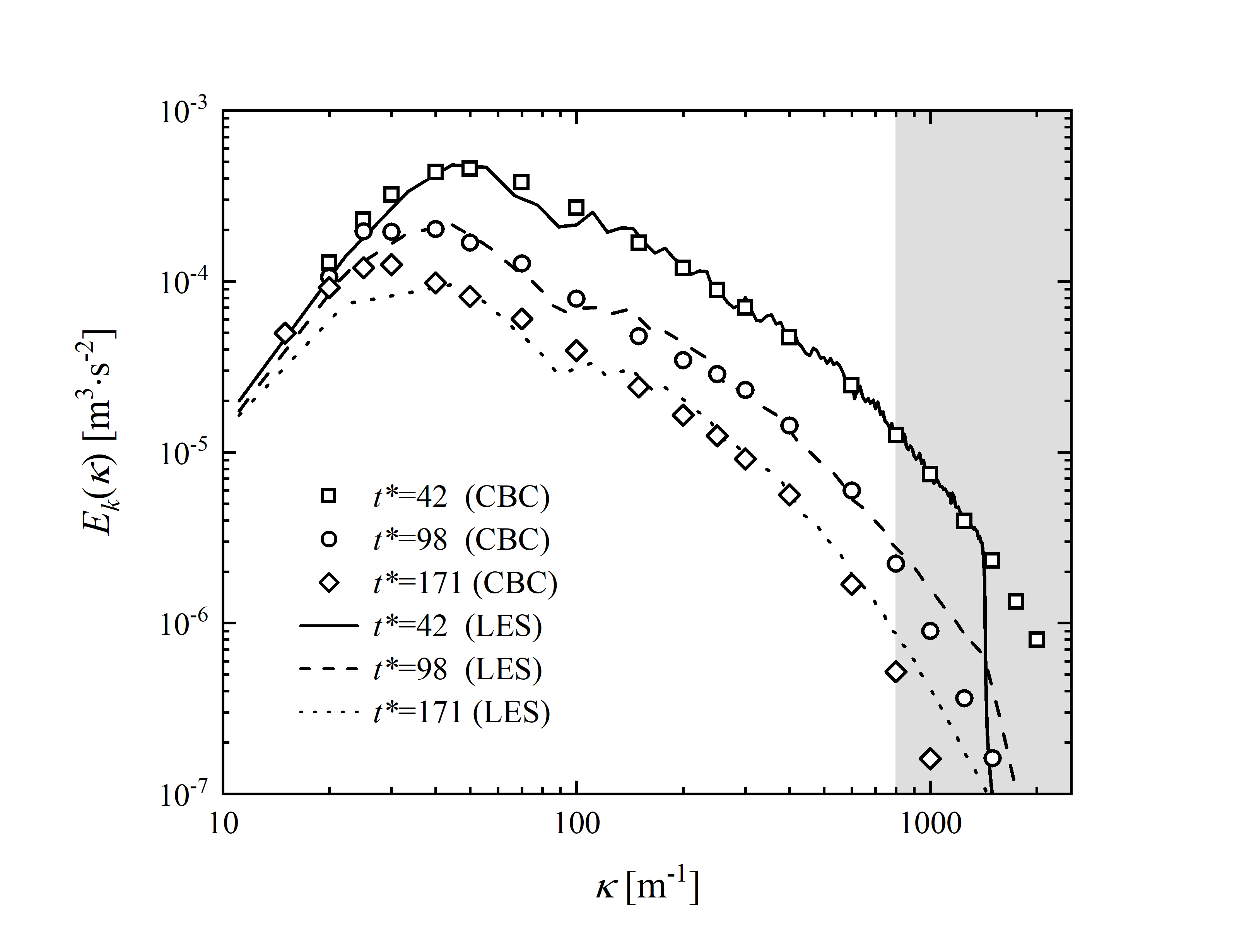}}
	\caption{Energy spectrum of homogeneous and isotropic turbulence at different moments during decay. Experimental data were obtained from the literature \citep{CBC:1971}.}
	\label{fig:HIspec}
\end{figure}

Case 1 covers a benchmark of decay of homogeneous and isotropic turbulence, where the size of the computational domain is $2\pi \times 9 \mathrm{cm}$ and the number of grids is $256^3$. When generating homogeneous and isotropic turbulence, the shifter version method, inverter version method and the extended method without divergence-free correction are all equivalent to the original spectrum-based method, there is no difference in the generated initial field. Large eddy simulations are carried out based on OpenFOAM, a finite volume method open source CFD code. A second-order Euler scheme is used for the time advancing, and spatial discretization employs the central difference scheme. The energy spectra of computation results at different time instants are compared with the experimental data, as shown in Fig.\ref{fig:HIspec}. The energy spectrum of the simulation agrees well with the CBC data, indicating that both the method implementation and the CFD code itself are valid.

\begin{figure}
	\centering
  \fbox{\includegraphics[width=8cm]{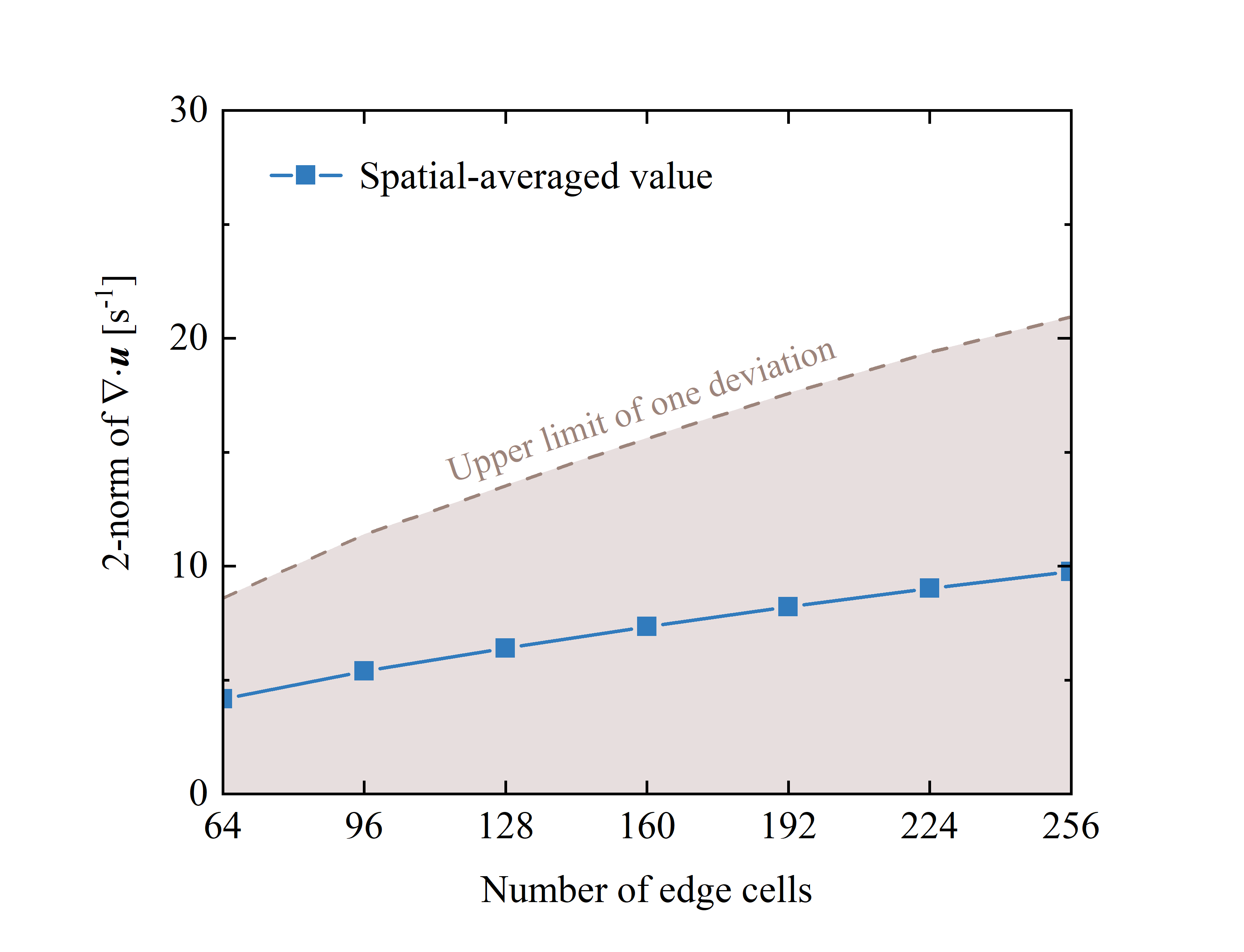}}
	\caption{Reference value indicating divergence-free under different mesh resolutions (obtained by homogenous and anisotropic turbulence generation computations).}
	\label{fig:bench}
\end{figure}

Although the divergence is strictly zero in theory, the numerical divergence of practical computation after discretization has a certain value due to the influence of a series of factors such as floating point number accuracy, numerical error, flow type, and divergence calculation method. Previous studies have shown that when the number of grids is sufficient, high-order interpolation methods have limited improvement in the accuracy of divergence calculation \citep{Yu:2014}. Therefore, the second-order central difference method is used to calculate divergence in all test cases. There is no doubt that the solenoidal property of the original spectral method handling inhomogeneity and isotropy is recognized. Therefore, we take the realistic divergence value of the proposed method producing homogeneous and isotropic turbulence as the reference or benchmark indicating divergence-free. This treatment eliminates other irrelevant factors in the subsequent discussion of the divergence-free error of the same type of box geometry. Fig.\ref{fig:bench} shows the actual divergence magnitude level $|u_{i,i}|$ of homogenous and isotropic turbulence at different grid resolutions. The blue line in the figure represents the spatial average of divergence magnitude $E$ (sample expectation), and the brown-gray area represents the range of $[\max(0,E-D),D]$, where $D$ is the sample standard deviation of the divergence magnitude distribution obtained based on spatial averaging as well.

\subsection{Anisotropic turbulence}
\label{ssec:rA}

Case 2 adopts the same box geometry as case 1 and tests four types of anisotropic turbulence while turbulent kinetic energy remains the same. Detailed components information is presented in Table \ref{tab:anisotype}. In the table, $R_0 = k/6 = \langle u_i u_i \rangle / 12$. The off-diagonal elements of the Reynolds stress of type A, type B, and type C are all 0, and the three diagonal elements are three principal stresses respectively. The tensor of type D contains non-zero off-diagonal components. Type A represents the anisotropy caused by one principal stress being significantly large, type B represents the anisotropy caused by one principal stress being equal to the average stress and the other two principal stresses being larger or smaller, and type C represents the anisotropy caused by one principal stress being smaller than the other two. Because type D contains non-zero off-diagonal elements, it means that the coordinate axis is not the same as the stress principal axis, which is a more common situation. The magnitude relationship between each component is designed to satisfy the stress magnitude relationship in typical channel flow (i.e., case 4): $\langle u_1 u_1 \rangle$ is the maximum, $\langle u_3 u_3 \rangle$ slightly greater than $\langle u_2 u_2 \rangle$, $\langle u_1 u_2 \rangle$ is negative, and the remaining components are all 0.

\begin{table}
  \caption{Four types of tested anisotropy.}
  \centering
  \begin{tabular}{cccc}
    \toprule
    \multicolumn{4}{c}{Type A}    \\
    \midrule
    $\langle u_i u_j \rangle / R_0$
            & $j = 1$ & $j = 2$ & $j = 3$ \\
    $i = 1$ & 10      & 0       & 0       \\
    $i = 2$ & 0       & 1       & 0       \\
    $i = 3$ & 0       & 0       & 1       \\
    \midrule
    \multicolumn{4}{c}{Type B}            \\
    \midrule
    $\langle u_i u_j \rangle / R_0$
            & $j = 1$ & $j = 2$ & $j = 3$ \\
    $i = 1$ & 7       & 0       & 0       \\
    $i = 2$ & 0       & 4       & 0       \\
    $i = 3$ & 0       & 0       & 1       \\
    \midrule
    \multicolumn{4}{c}{Type C}            \\
    \midrule
    $\langle u_i u_j \rangle / R_0$
            & $j = 1$ & $j = 2$ & $j = 3$ \\
    $i = 1$ & 5       & 0       & 0       \\
    $i = 2$ & 0       & 5       & 0       \\
    $i = 3$ & 0       & 0       & 2       \\
    \midrule
    \multicolumn{4}{c}{Type D}            \\
    \midrule
    $\langle u_i u_j \rangle / R_0$
            & $j = 1$ & $j = 2$ & $j = 3$ \\
    $i = 1$ & 8       & -2      & 0       \\
    $i = 2$ & -2      & 1       & 0       \\
    $i = 3$ & 0       & 0       & 3       \\
    \bottomrule
  \end{tabular}
  \label{tab:anisotype}
\end{table}

The shifter version and inverter version methods were used to generate these four types of anisotropic turbulence, respectively, and the corresponding statistical results based on spatial averaging are shown in Table \ref{tab:accuracy}. The data in the table are the turbulence field results obtained by only one realization (one sample), where the seeds of the random number engine in two versions are guaranteed to be the same. It is evident from the table that the inverter version results agree well with the given four anisotropies, basically ensuring accurate reproduction of the desired second-order correlation function. However, the Reynolds stress components depicted by the shifter version show a large deviation under all four anisotropies, leading to a tensor deformation. As mentioned above, the reason for this result is that the unit direction vector cannot maintain the original spherical distribution, and thus the covariance matrix deviates from the unit matrix.

\begin{table}
  \caption{The generation accuracy of desired anisotropic Reynolds stress by different versions of the proposed method.}
  \centering
  \begin{tabular}{ccc}
    \toprule
    Type A       & Shifter Ver. & Inverter Ver. \\
    \midrule
    $R_{11}/R_0$ & 8.13069      & 9.99790       \\
    $R_{22}/R_0$ & 1.89343      & 9.90726e-1    \\
    $R_{33}/R_0$ & 1.97590      & 1.01142       \\
    $R_{12}/R_0$ & 1.51089e-2   & -3.54073e-2   \\
    $R_{13}/R_0$ & -7.63861e-2  & -8.31628e-2   \\
    $R_{23}/R_0$ & 6.08866e-2   & 1.62758e-2    \\
    \midrule
    Type B       & Shifter Ver. & Inverter Ver. \\
    \midrule
    $R_{11}/R_0$ & 6.16861      & 7.04413       \\
    $R_{22}/R_0$ & 4.29450      & 3.92606       \\
    $R_{33}/R_0$ & 1.53690      & 1.02981       \\
    $R_{12}/R_0$ & 1.03419e-2   & 1.32772e-2    \\
    $R_{13}/R_0$ & -5.37689e-2  & -1.01037e-1   \\
    $R_{23}/R_0$ & 6.22320e-2   & 3.61277e-2    \\
    \midrule
    Type C       & Shifter Ver. & Inverter Ver. \\
    \midrule
    $R_{11}/R_0$ & 4.78214      & 5.03483       \\
    $R_{22}/R_0$ & 4.68151      & 4.93443       \\
    $R_{33}/R_0$ & 2.53635      & 2.03074       \\
    $R_{12}/R_0$ & 1.50719e-2   & 6.74909e-2    \\
    $R_{13}/R_0$ & -5.44608e-2  & -1.56510e-1   \\
    $R_{23}/R_0$ & 9.48676e-2   & 5.04841e-2    \\
    \midrule
    Type D       & Shifter Ver. & Inverter Ver. \\
    \midrule
    $R_{11}/R_0$ & 6.71562      & 7.90558       \\
    $R_{22}/R_0$ & 1.23883      & 1.00987       \\
    $R_{33}/R_0$ & 4.04554      & 3.08455       \\
    $R_{12}/R_0$ & -1.54632     & -2.00715      \\
    $R_{13}/R_0$ & -1.28067e-1  & -4.69213e-2   \\
    $R_{23}/R_0$ & 5.95098e-2   & 2.44947e-2    \\
    \bottomrule
  \end{tabular}
  \label{tab:accuracy}
\end{table}

Further investigating the results in Table \ref{tab:accuracy}, we noticed that the correlation deformation of the shifter version method exhibits a common pattern which we call anisotropy degeneration: any anisotropic tensor tends to degenerate toward an isotropic one $\delta_{ij}$. In the deformed tensors of type A, type B and type C, this pattern is expressed as three principal stress being more uniform: the larger principal stress becomes smaller and the smaller principal stress becomes larger. In the results of type D, this isotropy trend means that the off-diagonal elements are closer to 0. This anisotropic degradation pattern is positive in a way because it does not introduce errors or distortions that violate physics. Although the anisotropy intensity of the resultant turbulent field is weakened, it can be quickly restored to the real anisotropy state after development (calculation). Moreover, it is important to note that the anisotropic degradation presented in the shifter version described above does not necessarily apply to other spectrum-based methods that adopt similar ideas but different vector normalization algorithms, for example \citep{Kraichnan:1970,Smirnov:2001}. This is because the change in statistical distribution due to normalization at different steps is usually different. For example, when the three directions of a vector follow three independent Gaussian distributions, its covariance matrix is the identity matrix. However, after normalizing it to a unit vector, each component has a normal distribution of range $[-1,1]$ together with a 1/3 variance, and the three components are no longer independent. Although three times the covariance matrix of the latter is still the identity matrix, the two random vectors are already different from a statistical view. After an identical tensor operation, e.g. tensor multiplication, the distribution of these two vectors will be significantly different (even the covariance matrix is not the same anymore). We tried to only use Gaussian distribution to generate vector components and did not perform unit vector normalization in the intermediate process, but just took an operation similar to dividing by 3 to ensure that the covariance matrix of the vector remained unchanged. As a result, the Reynolds stress distortion obtained by this treatment was even worse than that of the shifter version and there was no similar pattern of anisotropy degeneration. A very unreasonable oversize and undersize were observed. However, since this treatment does not carry out normalization operations, the analytical formula of statistics can be obtained in theoretical analysis. Among them, we get two physical quantities characterizing the degree of anisotropy:

\begin{align}
  \label{eq:aniso:d}
  A_d &=
    \frac{
      \left(R_{11}-R_{22}\right)^2
    + \left(R_{11}-R_{33}\right)^2
    + \left(R_{22}-R_{33}\right)^2
    }{
      \left(R_{11} + R_{22} + R_{33}\right)^2
    },  \\
  \label{eq:aniso:n}
  A_n &=
    \frac{
      R_{12}^2 + R_{13}^2 + R_{23}^2
    }{
      \left(R_{11} + R_{22} + R_{33}\right)^2
    }.
\end{align}

In the equation, $A_d$ denotes the anisotropy of the diagonal elements of the matrix and $A_n$ denotes the anisotropy of the off-diagonal elements.

\begin{figure}
	\centering
  \fbox{
    \includegraphics[width=8cm]{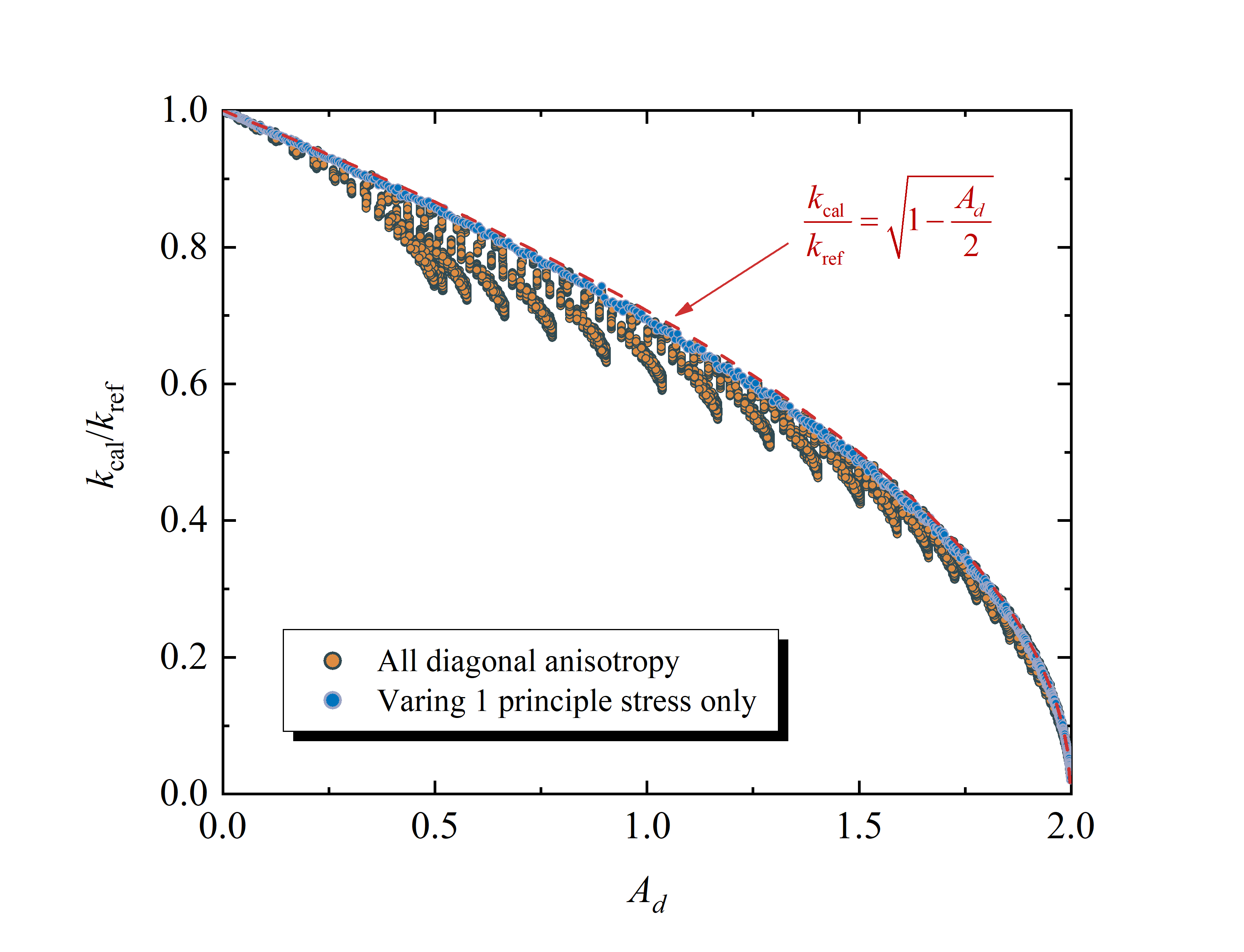}
    \includegraphics[width=8cm]{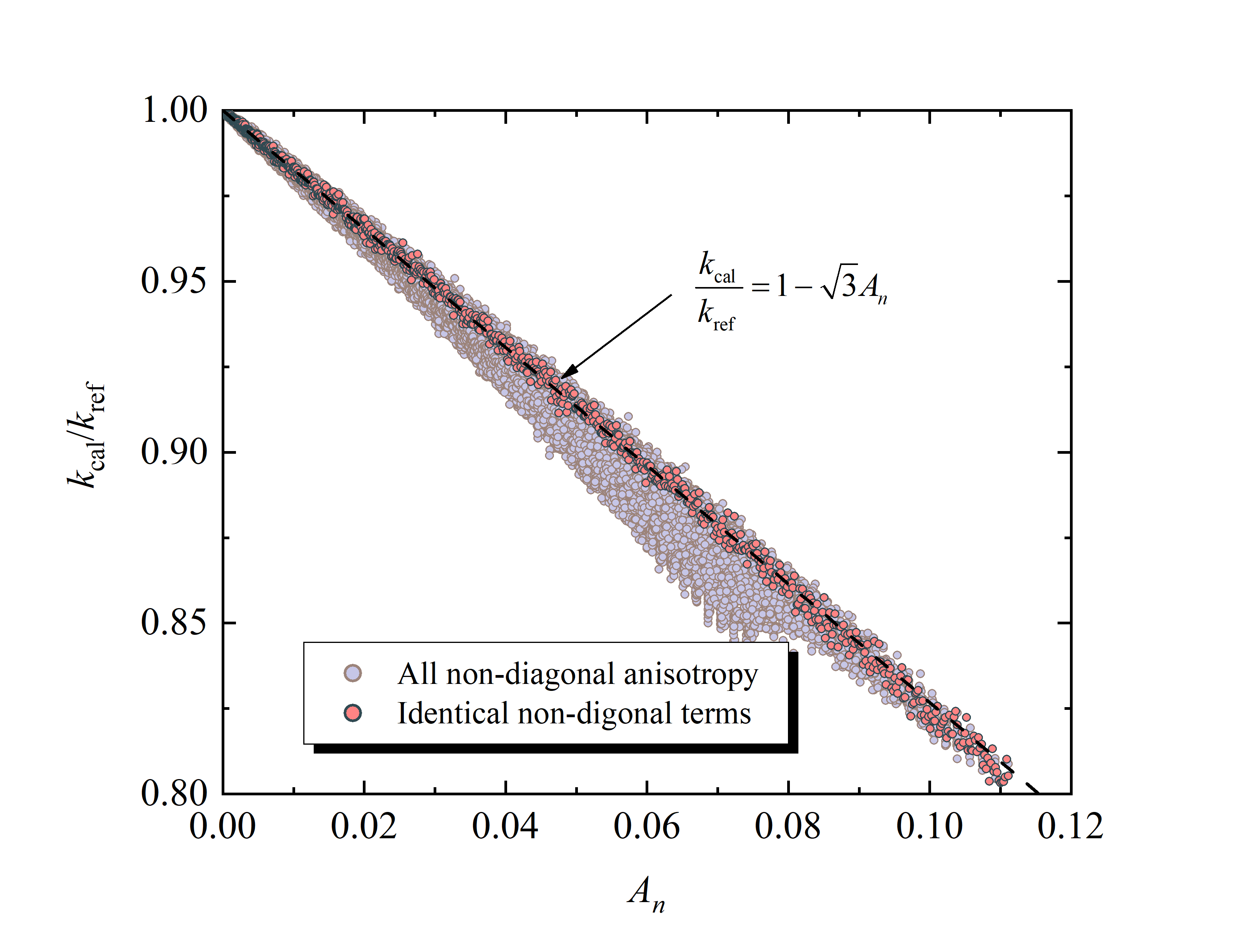}
  }
	\caption{The reduction in turbulent kinetic energy due to the mapped direction vector. (\textit{a}) diagonal element anisotropy, and (\textit{b}) off-diagonal element anisotropy.}
	\label{fig:anideg}
\end{figure}

In addition to the anisotropic degradation, the shifter version method leads to a decay of the turbulent kinetic energy, which is not present in the inverter version as well. This TKE reduction is also caused by the change in the statistical distribution of the unit direction vector $\sigma_i^{(n)}$. As mentioned above, it is very difficult and tedious to theoretically derive the exact distribution form and covariance matrix because the intermediate step of the shifter version correction method introduces vector normalization operations. Therefore, we used numerical experiments to test the behavior of this turbulent kinetic energy decay with respect to anisotropy intensity $A_d$ and $A_n$, and the relevant results are shown in Fig.\ref{fig:anideg}. Each data point in Fig.\ref{fig:anideg} is calculated by averaging 20000 samples. Fig.\ref{fig:anideg}(\textit{a}) and Fig.\ref{fig:anideg}(\text{b}) show how the reduction of TKE varies with the diagonal anisotropy intensity $A_d$ and the off-diagonal anisotropy intensity $A_n$, respectively. The off-diagonal components of the input stress tensor in Fig.\ref{fig:anideg}(\textit{a}) are 0, and each diagonal one varies range from 1 to 100 by 2 orders of magnitude, respectively. It can be clearly seen from the figure that at this time, different types of anisotropy do not completely fall into a single curve, but present an obvious distribution area. The range of this distribution area is large at $A_d = 0.5--1.5$, and tends to an overlapping curve when the anisotropy is large and small. The blue dot set in Fig.\ref{fig:anideg}(\textit{a}) shows the influence of type A anisotropy, a certain principal stress is significantly larger, on the TKE decay. This is also the case that the diagonal anisotropy intensity is close to its upper limit of 2. It can be seen that this type basically determines the upper limit of the turbulent kinetic energy decay, which coincides with the analytical curve of $k_{\mathrm{cal}} = k_{\mathrm{ref}} \sqrt{1 - 0.5 A_d}$. It can be found that the anisotropy of the principal stress has a great influence on turbulent kinetic energy reduction. When the $A_d$ tends to 2, the turbulent kinetic energy will even be close to 0. In Fig.\ref{fig:anideg}(\textit{b}), the diagonal components are all kept as 1, and the off-diagonal ones are varied from 0 to 1/3 to ensure that the matrix is always positive definite. It can be found that the influence of off-diagonal anisotropy also has a law similar to that of diagonal anisotropy, i.e., it does not coincide with a single curve, but its influence is significantly smaller than that of the diagonal one. Even when the three off-diagonal elements reach the maximum value, the calculated turbulent kinetic energy reduction is less than 20\%. The red dot in the figure shows the case when the three off-diagonal elements are equal, and again the upper limit of the decay is basically determined, which is satisfied by the analytical curve $k_{\mathrm{cal}} = k_{\mathrm{ref}} \left( 1 - \sqrt{3} A_n \right)$.

Therefore, the shifter version method exhibits two unpleasant characteristics: anisotropic degradation and TKE reduction. The latter can be scaled to recover according to $A_d$ and $A_n$ by the analytical function, while the former has no effective correction strategy if keeping the framework of the shifter version. But degenerated anisotropic can quickly recover to the real state in the actual calculation, as shown in the applications of using isotropic turbulence as the initial condition for anisotropic computations. On the other hand, the inverter version does not have the above two issues at all and performs well in reproducing the desired correlations and statistics, so the following divergence-free analysis is mainly done with the inverter version (although the shifter version also has similar performance in the divergence-free behavior).

\begin{figure}
	\centering
  \fbox{
    \includegraphics[width=8cm]{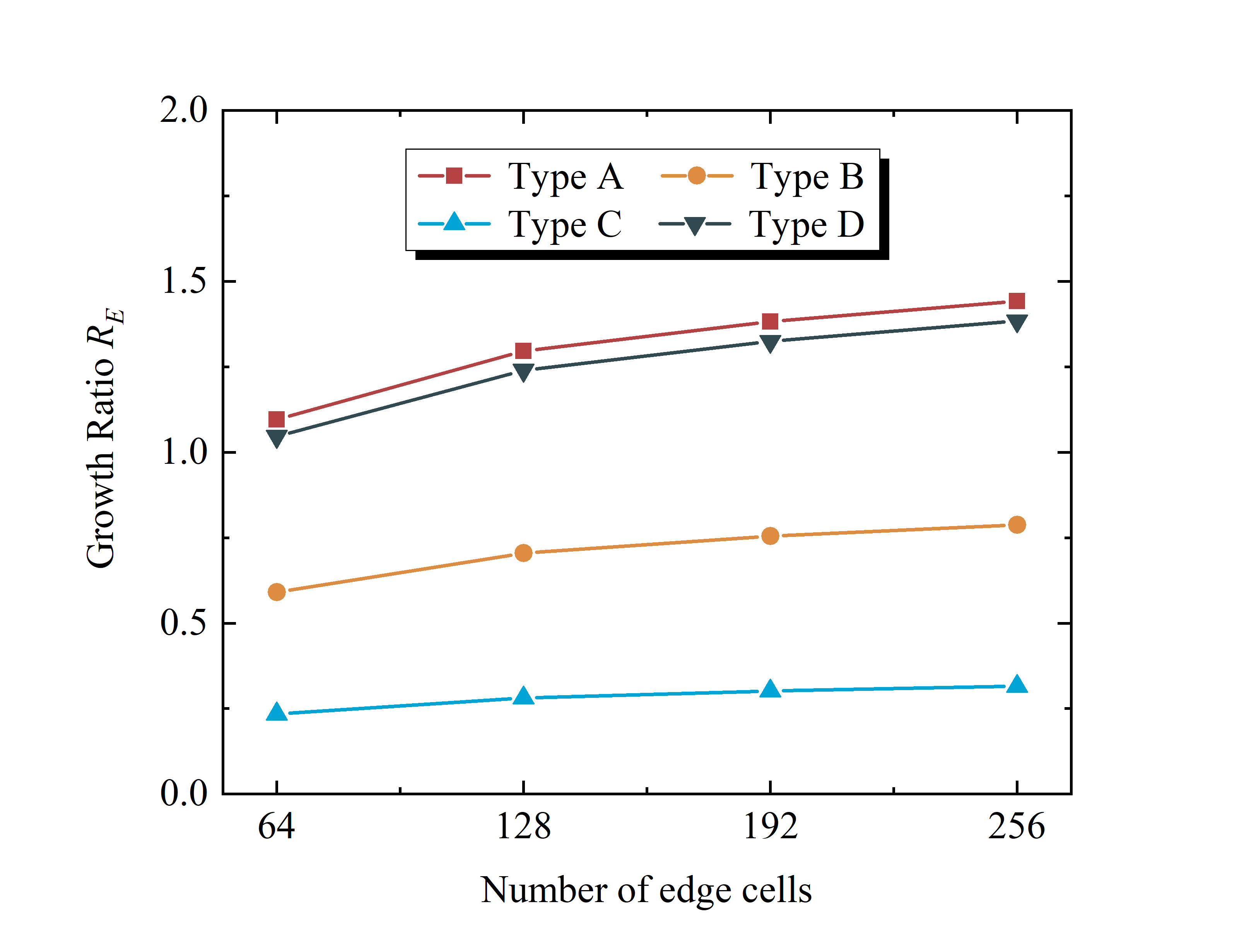}
    \includegraphics[width=8cm]{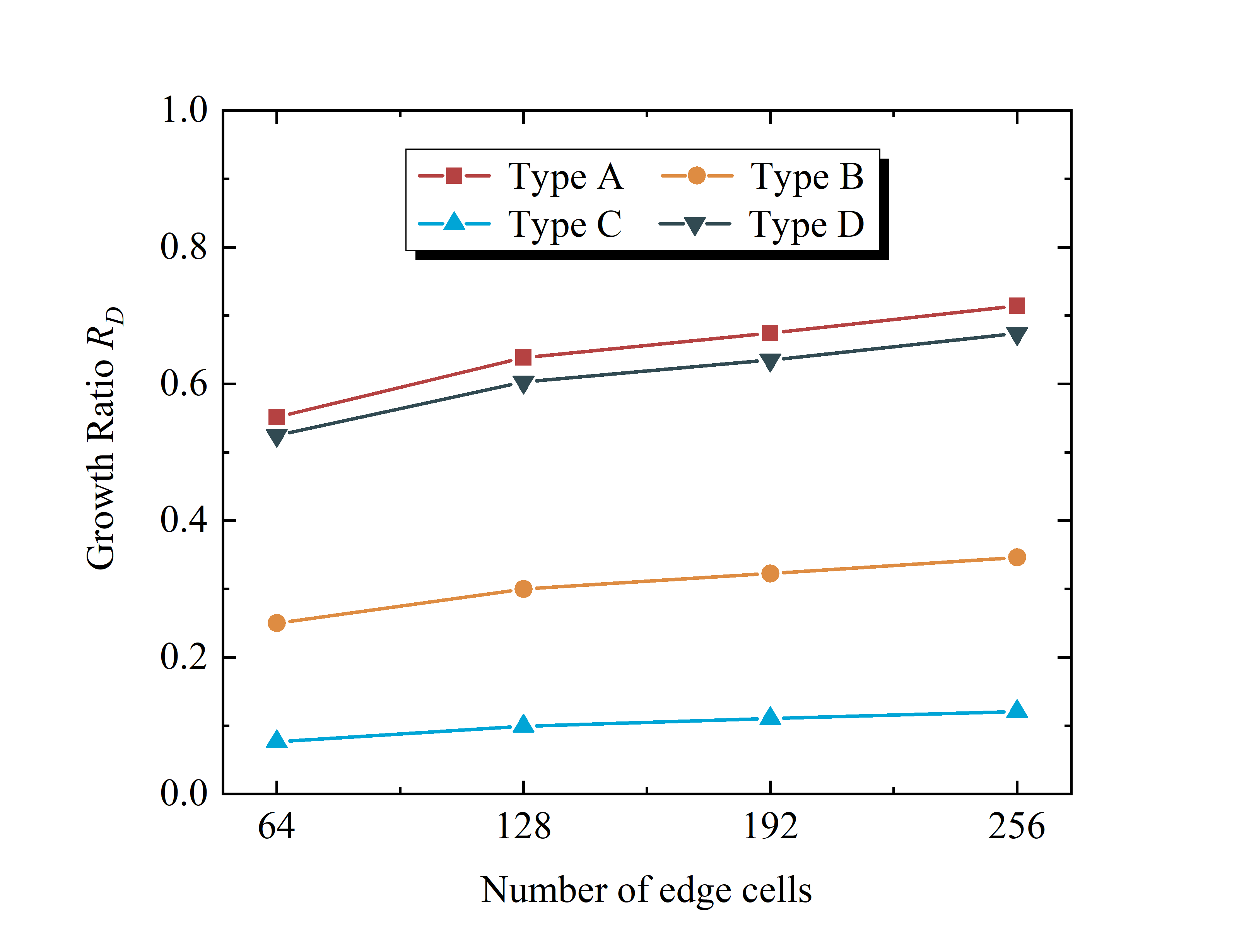}
  }
	\caption{Variation of the divergence-free error due to different anisotropy types with respect to increasing mesh resolution. (\textit{a}) mean value/sample expectation, (\textit{b}) deviation value/sample standard deviation.}
	\label{fig:divErr}
\end{figure}

Fig.\ref{fig:divErr} shows the increase of the divergence-free error if the divergence correction is not applied (i.e. the extended method in Section \ref{ssec:extent}). The error growth factor in the figure is defined as follows:
\begin{equation}
  R_X = \frac{X_{\mathrm{uncorrected}}-X_{\mathrm{standard}}}{X_{\mathrm{standard}}},
\end{equation}
where, $X$ is the statistical indicator used, in Fig.\ref{fig:divErr}(\textit{a}) is the mean value $E$, and in Fig.\ref{fig:divErr}(\textit{b}) is the deviation $D$. $X_{\mathrm{standard}}$ represents the zero divergence benchmark under the same mesh resolution, which is the result of the calculation in Section \ref{ssec:rHI} (Fig.\ref{fig:bench}). Four grid numbers of $64^3$, $128^3$, $192^3$ and $256^3$ are adopted. It is obvious from the figure that the more anisotropic the turbulent the larger the divergence-free error, especially for type A and type D. The behavior of the anisotropy intensity relationship is consistent with the relative magnitudes of the $A_d$ and $A_n$ calculated by Eq.\eqref{eq:aniso:d} and Eq.\eqref{eq:aniso:n}. In addition, the relative growth rate of error $R_X$ also showed a mild dependence on mesh resolution, i.e., the larger the number of grids, the higher the relative growth rate of error. The behavior of the mean value $E$ in the figure is basically the same as that of the deviation value $D$, which means the observation is independent of the chosen indicator.

\begin{figure}
	\centering
  \fbox{
    \includegraphics[width=8cm]{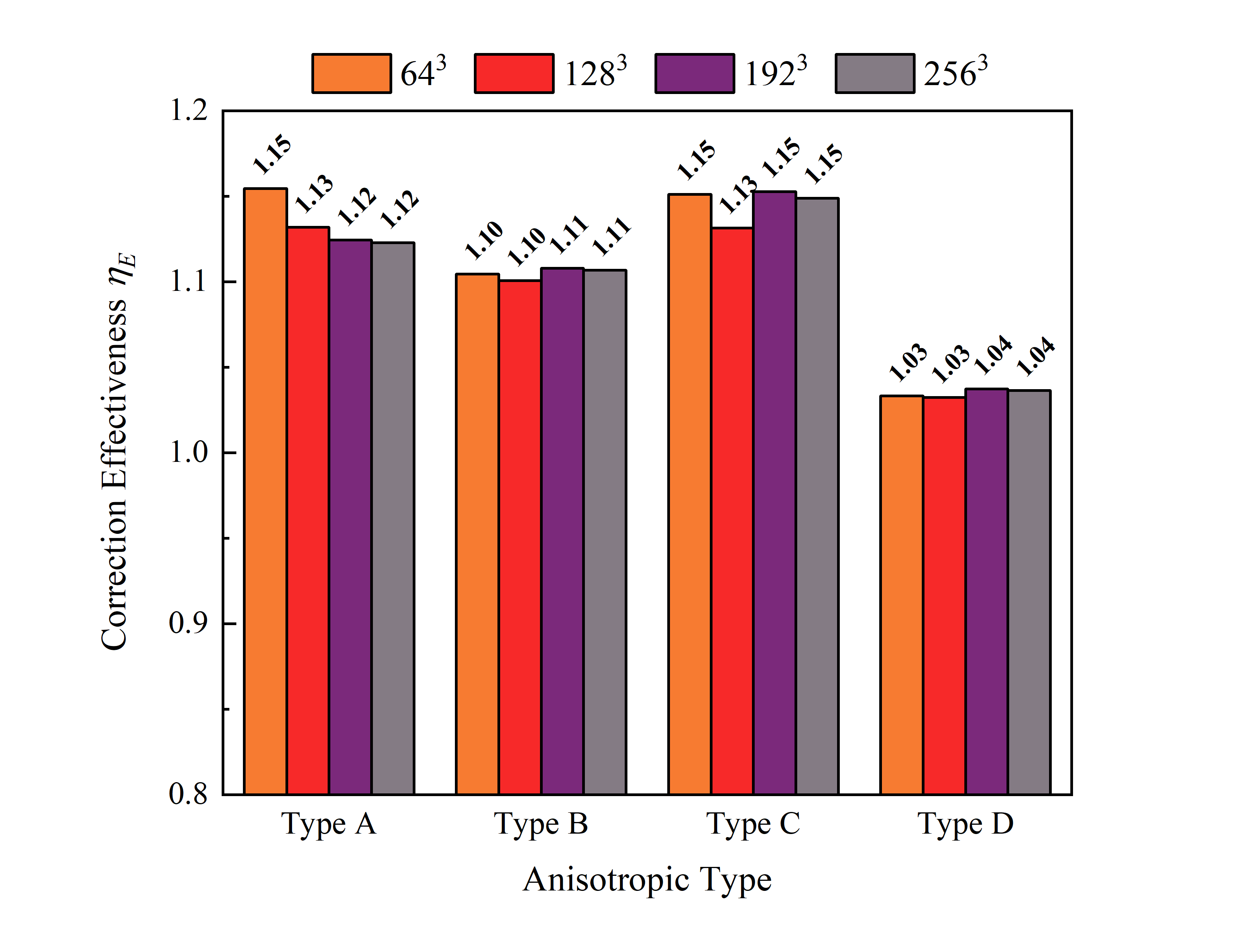}
    \includegraphics[width=8cm]{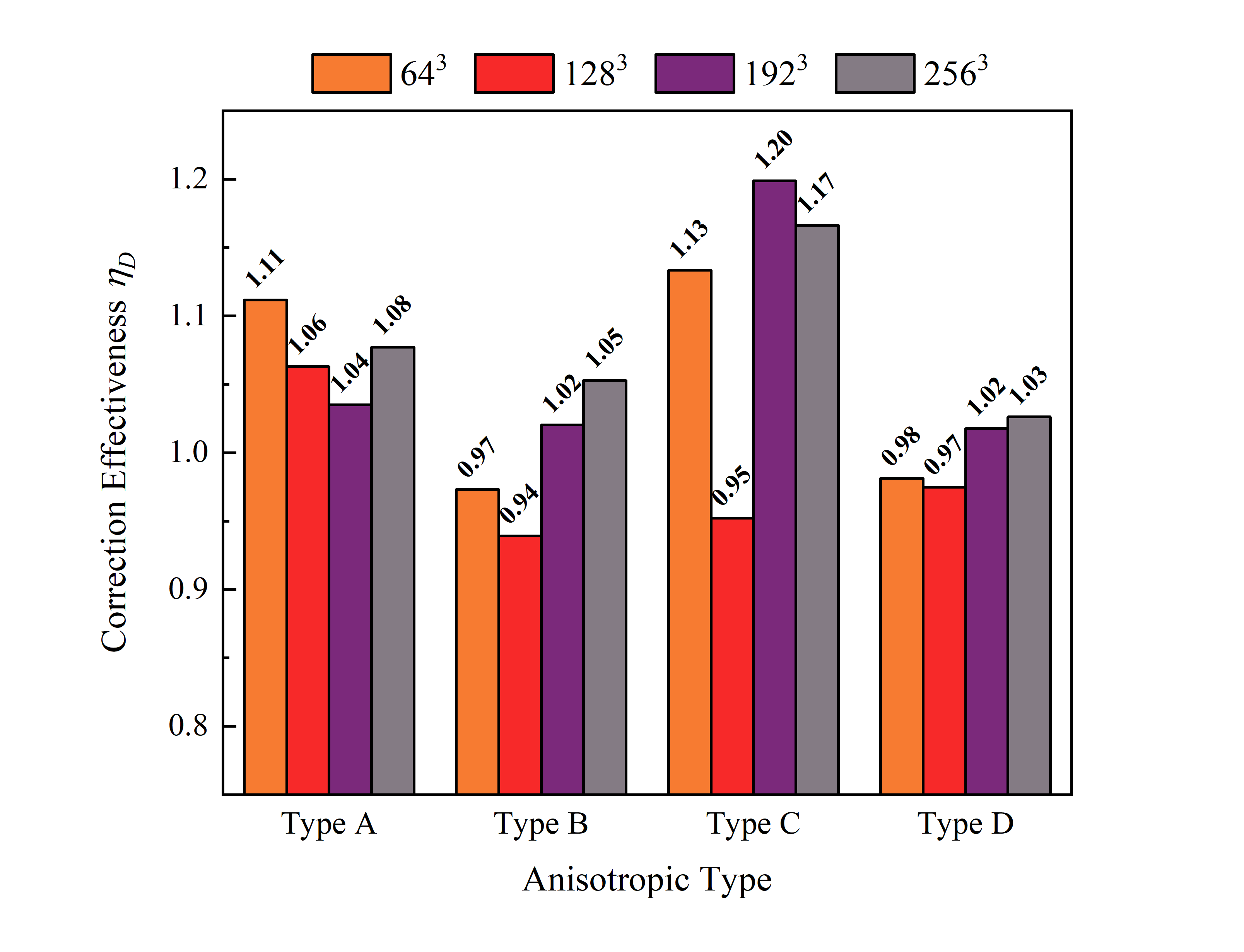}
  }
	\caption{Effectiveness of the inverter version method on the correction of the divergence-free error for different anisotropy types. (\textit{a}) mean value/sample expectation, (\textit{a}) deviation value/sample standard deviation.}
	\label{fig:effect}
\end{figure}

Fig.\ref{fig:effect} illustrates the effectiveness of the inverter version spectrum-based method for the correction of divergence growth. The correction effectiveness in the figure is calculated using the following formula:
\begin{equation}
  R_X =
    \frac{
      X_{\mathrm{corrected}} - X_{\mathrm{standard}}
    }{
      X_{\mathrm{uncorrected}} - X_{\mathrm{standard}}
    },
\end{equation}
where, $X_{\mathrm{corrected}}$ is the calculation result of the divergence-free method. When $\eta_X$ is close to 0, it means that the method does not correct the error at all. As $\eta_X$ approaches 1, it indicates that the method basically corrects 100\% of the increase in divergence caused by anisotropy. Due to the existence of factors such as numerical errors, the correction effectiveness calculated in this way may exceed one. Fig.\ref{fig:effect}(\textit{a}) and (\textit{b}) present the effectiveness represented by the mean and deviation values, respectively. It can be seen from the figure that the inverter version method basically corrects the four types of anisotropy close to 100\%, and the minimum value in Fig.\ref{fig:effect}(\textit{a}) also reaches 1. Due to the characteristics of the standard deviation itself, the behavior of the correction performance in Fig.\ref{fig:effect}(\textit{b}) is different from that in Fig.\ref{fig:effect}(\textit{a}), but the minimum value of effectiveness is also above 90\%, which fully illustrates the qualification of the inverter version method in correcting the divergence-free error caused by anisotropy. Moreover, even though a mesh dependence is observed in uncorrected errors shown in Fig.\ref{fig:divErr}, the behavior in Fig.\ref{fig:effect} is independent of the grid size, which indicates that the inverter method is grid independent for the correction of the divergence-free error.

\subsection{Inhomogeneous turbulence}
\label{ssec:rI}

Case 3 also uses box geometry, but in order to study performance processing inhomogeneous turbulence, an artificial distribution of correlations along the $x$ direction is set as follows:
\begin{equation}
\label{eq:inhomo}
  \langle u_i u_j \rangle =
    \langle u_i u_j \rangle_{\mathrm{mean}}
    \left[ 1 - A_0 \cos \left( 2\pi\frac{x}{x_d} \right) \right],
\end{equation}
where $x_d$ is the total length of the box domain in the $x$ direction. The $y$ and $z$ directions are still set to maintain homogeneity so that in addition to ensemble averaging for multiple realizations, the planar averaging operation can still be performed to study the performance of the new method in a single realization. According to the Fourier series in Eq.\eqref{eq:ext:v}, it can be seen that the minimum wave number of the generated velocity $u_i$ is the $2\pi/x_d$. Therefore, the minimum wave number of $u_i u_j$ is $\pi/x_d$, i.e. the maximum wavelength identified is $0.5x_d$. The Fourier mode of wavelength $x_d$ in Eq.\eqref{eq:inhomo} is clearly not in the recognizable range and therefore, as mentioned above, is a good choice for characterizing macroscopic inhomogeneity. By controlling the parameter $A_0$ ($0 \leq A_0 \geq 1$), distributed turbulence with different inhomogeneities can be obtained. The calculation results in this section are all acquired employing the inverter version.

\begin{figure}
	\centering
  \fbox{
    \includegraphics[width=8cm]{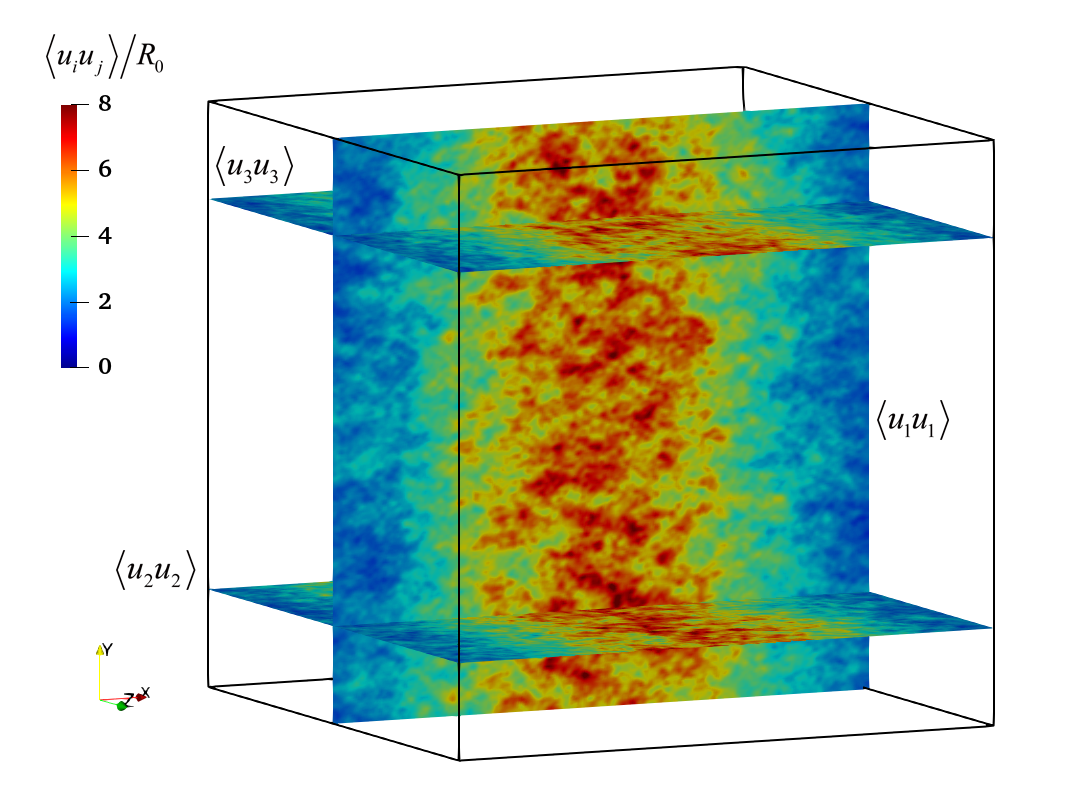}
    \includegraphics[width=8cm]{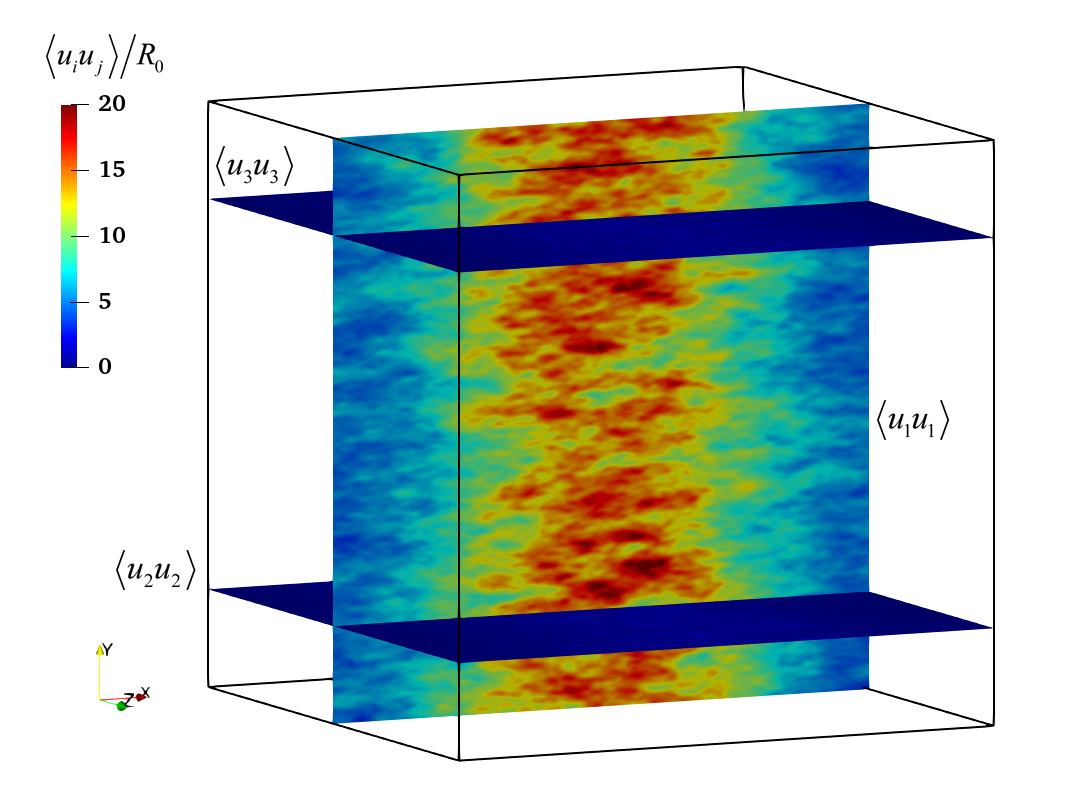}
  }
	\caption{Spatial distribution of statistical results of Reynolds stress components (100 realizations of turbulence generation). (\textit{a}) isotropic turbulence, (\textit{b}) anisotropic turbulence (type A).}
	\label{fig:ic}
\end{figure}

Under the condition that the spatial-averaged turbulent kinetic energy is the same, the isotropic but inhomogeneous turbulence and the type A anisotropic turbulence satisfying Eq.\eqref{eq:inhomo} are generated respectively. The spatial distribution of stresses after 100 consecutive sampling of generation and ensemble averaging is presented in Fig.\ref{fig:ic}. The $z$-crosssection in the figure shows the principal stress in the $x$-direction, and the two $y$-crosssections show the distribution of the other two principal stresses respectively. It is observed in the figure that both isotropic and anisotropic turbulence accurately restores the inhomogeneous distribution of the cosine modes. Compared with the situation where the three principal stress distributions in Fig.\ref{fig:ic}(\textit{a}) are approximately the same, a significantly larger value of components $\langle u_1 u_1 \rangle$ is observed in Fig.\ref{fig:ic}(\textit{b}), indicating that the method also has good accuracy in reproducing inhomogeneous anisotropy.

\begin{figure}
	\centering
  \fbox{
    \includegraphics[width=8cm]{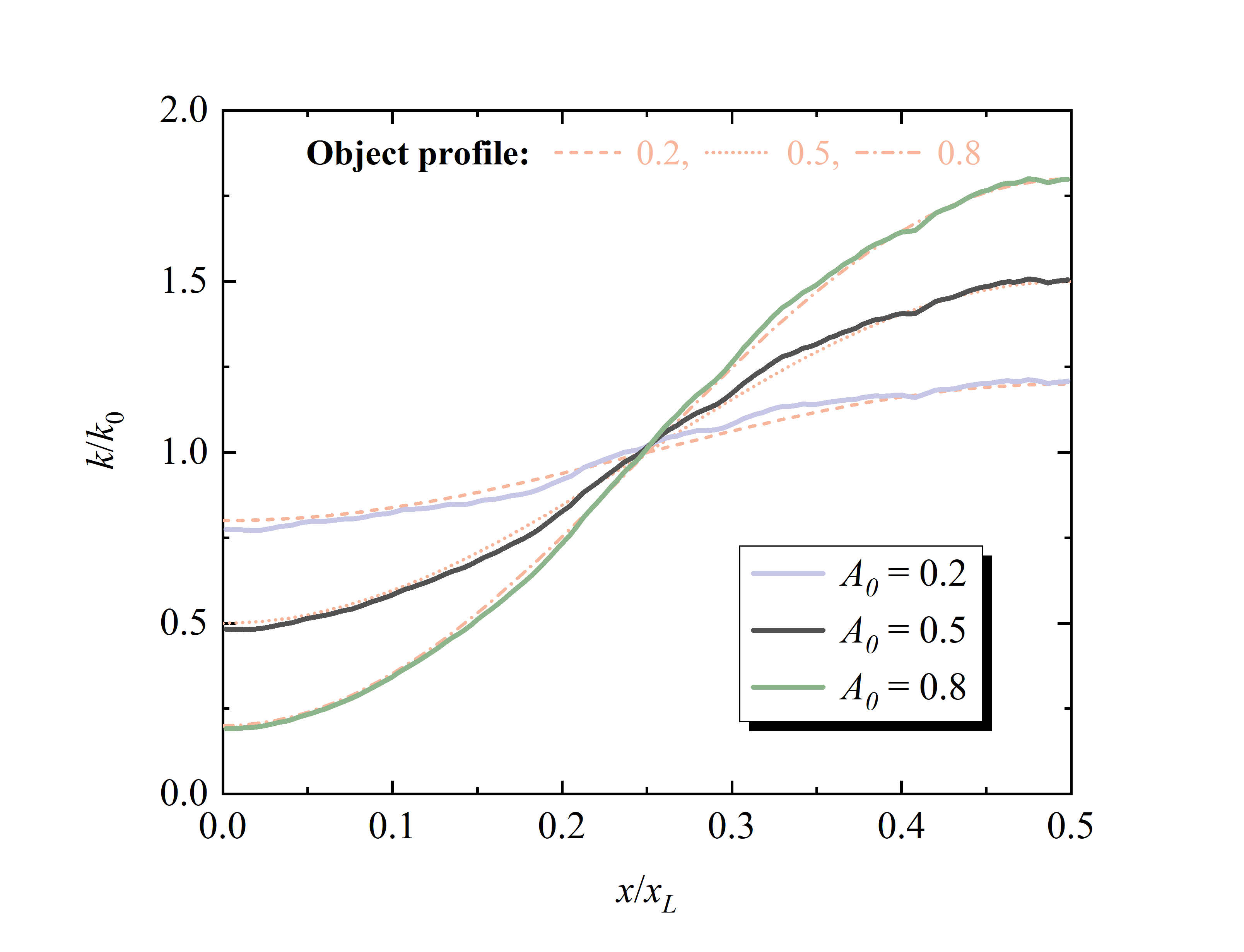}
    \includegraphics[width=8cm]{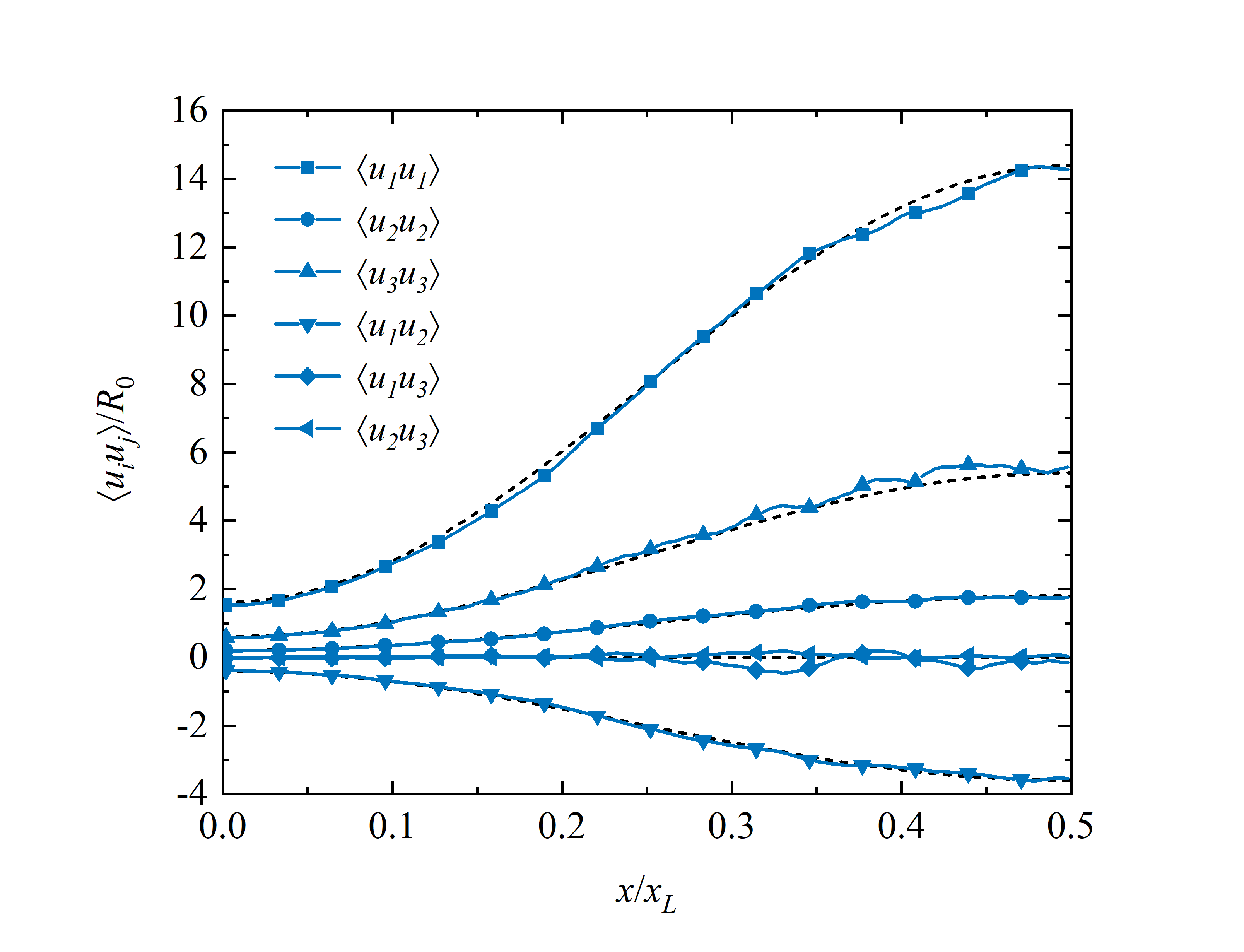}
  }
	\caption{The divergence-free method reproduction of the target inhomogeneous distribution of turbulence correlations (single realization, $y$-$z$ plane averaging). (\textit{a}) turbulent kinetic energy $k$ for different inhomogeneities, and (\textit{b}) Reynolds stress component of the type D anisotropy for $A_0 = 0.8$.}
	\label{fig:ihomokR}
\end{figure}

In order to further verify the reduction effect of the proposed method on the inhomogeneity, distributions of the turbulent kinetic energy and Reynolds stress components in Fig.\ref{fig:ihomokR} are obtained by a single realization and cross-plane averaging. Fig.\ref{fig:ihomokR}(\textit{a}) shows the TKE distribution for different inhomogeneity intensities reflected by $A_0$ magnitude. It is evident that the new method has good accuracy for all distributions of $k$, even for a single realization. Fig.\ref{fig:ihomokR}(\textit{b}) shows the verification results of the anisotropy of type D. Obviously, the reproduction characterization of all distributions is in good agreement with either the three normal stresses or the one non-zero shear stress. The distribution of $\langle u_1 u_3 \rangle$ and $\langle u_2 u_3 \rangle$ basically stays at zero with mild fluctuations. The larger fluctuations in the components $\langle u_1 u_3 \rangle$ are due to larger velocities in the two associated directions.

To observe the absolute divergence increase caused by anisotropy and inhomogeneity in both anisotropic and inhomogeneous turbulence, the special-averaged divergence levels of isotropy, type A anisotropy and type D anisotropy at $A_0 = 0$ (homogeneity), 0.2, 0.5 and 0.8 were computed. Relevant results are shown in Table \ref{tab:iu}. Turbulence intensity varies at different spatial locations due to inhomogeneity. Therefore, in order to eliminate the influence of this factor on statistics, averaged results in the table are dimensionless normalized divergence $|u_{i,i}|/u_t$.

\begin{table}
  \caption{Divergence-free errors (uncorrected) of synthesized turbulence with different inhomogeneity intensities.}
  \centering
  \begin{tabular}{ccccccc}
    \toprule
          & \multicolumn{2}{c}{Isotropy}
          & \multicolumn{2}{c}{Type A anisotropy}
          & \multicolumn{2}{c}{Type D anisotropy} \\
    \cmidrule{2-7}
    $A_0$ & Mean    & Deviation & Mean    & Deviation & Mean    & Deviation  \\
    0     & 44.3570 & 50.8523   & 105.751 & 85.1091   & 108.345 & 87.1736    \\
    0.2   & 44.3546 & 50.8510   & 105.749 & 85.1060   & 108.342 & 87.1709    \\
    0.5   & 44.3360 & 50.8411   & 105.725 & 85.0870   & 108.320 & 87.1536    \\
    0.8   & 44.2757 & 50.8047   & 105.632 & 85.0154   & 108.240 & 87.0894    \\
    \bottomrule
  \end{tabular}
  \label{tab:iu}
\end{table}

The calculated results of the mean value and deviation value of the divergence level in the table show that the increase of inhomogeneity intensity has almost no effect on the increase of divergence level: when $A_0$ increasing from 0 to 0.8, the divergence level of both isotropic and anisotropic turbulence nearly remain the same. Take isotropic turbulence as an example, its mean value remains around 40, while the standard deviation remains almost unchanged at around 51 in all inhomogeneities. However, when the flow changes from isotropic to anisotropic, there is a large increase in the divergence-free error. For type A anisotropy, the mean value increases from 44 to 106, and the deviation increases from 51 to 85. For type D anisotropic turbulence, the mean value increased from 44 to 108 and the deviation increased from 51 to 87. The results in Table \ref{tab:iu} show that under the inhomogeneous distribution of this form (cosine distribution of wavelength  ), the increase in the error caused by anisotropy far exceeds the one caused by inhomogeneity. When $A_0 = 0.8$, where the most intensive inhomogeneity of this wave number is nearly approached, the increase in divergence-free errors brought inhomogeneity is still not obviously observed. No doubt increasing the wave number (frequency) of the mode can indeed bring a larger local gradient and inhomogeneity, but notice that this part of the mode can already be identified and constructed by spectrum-based methods. It is doubtful whether these high-frequency signals can be regarded as macroscopic inhomogeneity rather than small-scale turbulence.

Table \ref{tab:ic} shows the inverter version's correction performance of the two types of anisotropy in Table \ref{tab:iu} (isotropic data directly follows the results of Table \ref{tab:iu}). It can be seen from the table that both the mean and deviation value of the two types of anisotropic turbulence generated by the correction method reach the actual divergence level of isotropic turbulence under the same conditions, so it can be considered that the requirements of divergence-free are satisfied in practice. Although the inverter version correction does not account for inhomogeneity errors, the correlation reconstruction matrix varies in space, leading to effective correction for inhomogeneous anisotropy as well. This is reflected in Table \ref{tab:ic} that results of different intensities $A_0$ have similar correction effectiveness.

\begin{table}
  \caption{Divergence-free errors (after the inverter version correction) of synthesized turbulence with different inhomogeneity intensities.}
  \centering
  \begin{tabular}{ccccccc}
    \toprule
          & \multicolumn{2}{c}{Isotropy}
          & \multicolumn{2}{c}{Type A anisotropy}
          & \multicolumn{2}{c}{Type D anisotropy} \\
    \cmidrule{2-7}
    $A_0$ & Mean    & Deviation & Mean    & Deviation & Mean    & Deviation  \\
    0     & 44.3570 & 50.8523   & 42.1230 & 49.9550   & 36.5007 & 48.0495    \\
    0.2   & 44.3546 & 50.8510   & 42.1259 & 49.9564   & 36.5075 & 48.0469    \\
    0.5   & 44.3360 & 50.8411   & 42.1413 & 49.9588   & 36.5565 & 48.0499    \\
    0.8   & 44.2757 & 50.8047   & 42.2085 & 49.9608   & 36.7595 & 48.0733    \\
    \bottomrule
  \end{tabular}
  \label{tab:ic}
\end{table}

\subsection{Typical inhomogeneous and anisotropic turbulent flow}
\label{ssec:rAI}

The fourth case is a typical inhomogeneous and anisotropic turbulent flow common in practical engineering: channel flow with fully developed boundary layers. Because the first three cases have verified the method in correlation accuracy and the divergence-free property, case 4 is mainly used to test the performance of the proposed method (inverter version) in the practical application used for practical turbulent flow and the inhomogeneity influence on divergence errors is checked analytically for the channel flow. The periodic channel flow parameters and grids selected in this case are consistent with those in DNS literature \citep{Kim:1987}, and the other information can be found in \citep{Mansour:1988}. The LES in this case does not use any subgrid stress model due to the DNS-resolution mesh. (Complete LES calculations have been performed and the validity of the CFD code is not shown here due to content constraints and limited relevance.) The reference data used for comparison and the data used for theoretical analysis in this section are the calculation results of DNS in the literature.

\begin{figure}
	\centering
  \fbox{\includegraphics[width=8cm]{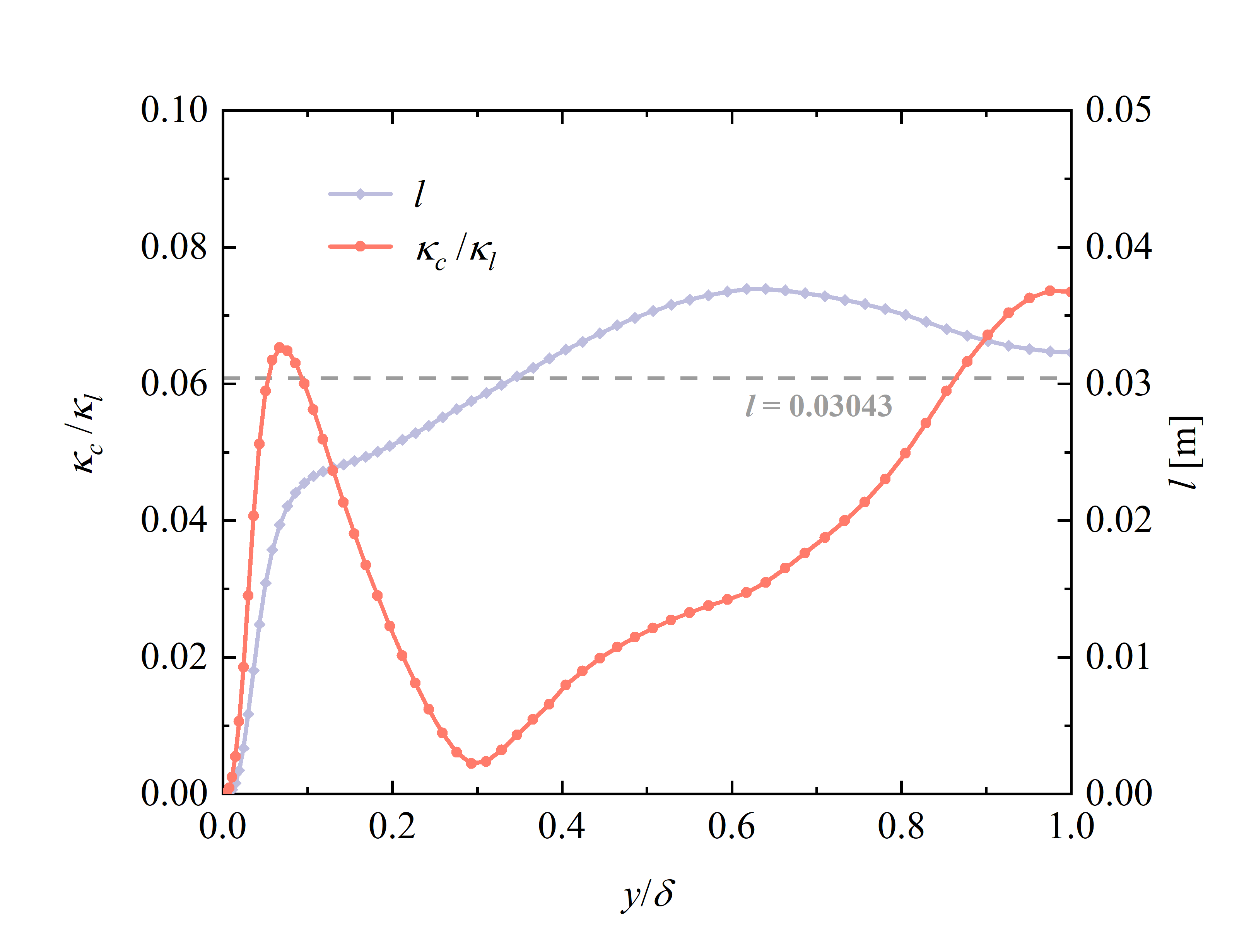}}
	\caption{The proportion of influence of inhomogeneity and the spatial distribution of turbulence integral scale in channel flow.}
	\label{fig:cf:l}
\end{figure}

Firstly, a dimensionless variable $\kappa_c/\kappa_l$ indicating the influence of inhomogeneity is calculated based on the data of channel flow and Eq.\eqref{eq:sig:kk1}. The distribution of this wave number ratio is shown in the red line in Fig.\ref{fig:cf:l}. It can be seen from the figure that the critical wave number of inhomogeneity $\kappa_c$ in the entire domain is below 8\% of the energy-containing wave number $\kappa_l$, and $\kappa_c$ remains below 4\% in most areas. This illustrates that in the channel flow, the error term $Er_4$ related to inhomogeneity is far less than $Er_5$, so the use of the inverter version correction method is fully qualified for the practical generation requirement of obtaining solenoidal turbulence.

In channel flows, the turbulent integral length scale (energy-containing scale) is often not a constant, but has a specific spatial distribution, as shown by the lilac line in Fig.\ref{fig:cf:l}. The integral scale of the region near the core is larger than that of the region near the wall. However, the non-uniform integral scale directly leads to different energy spectra in different spatial locations (multiple items in Eq.\eqref{eq:spectra} are directly or indirectly dependent on $l$ or turbulent dissipative rate), and as shown in Eq.\eqref{eq:errors}, it will bring the divergence-free error of $Er_1$ at the same time. To analyze the influence brought by the distributed integral scale, case 4 generates the inhomogeneous and anisotropic turbulence of the channel flow from the uniform integral scale of a single constant (the mean value represented by the gray dashed lines in Fig.\ref{fig:cf:l}) and the distributed integral scale profile respectively, and compares the differences in the statistics accuracy and divergence-free characteristic between the two.

\begin{figure}
	\centering
  \fbox{
    \includegraphics[width=8cm]{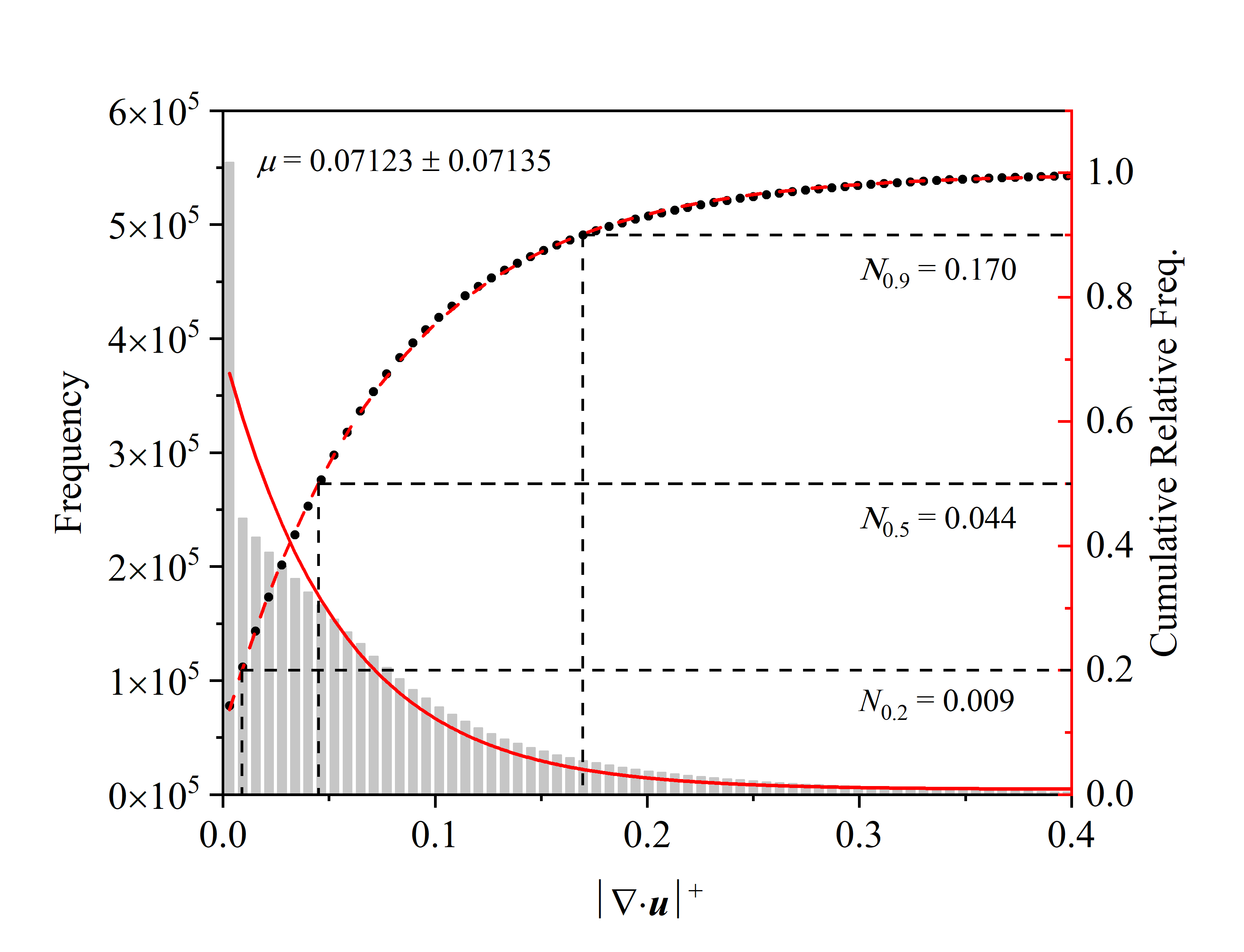}
    \includegraphics[width=8cm]{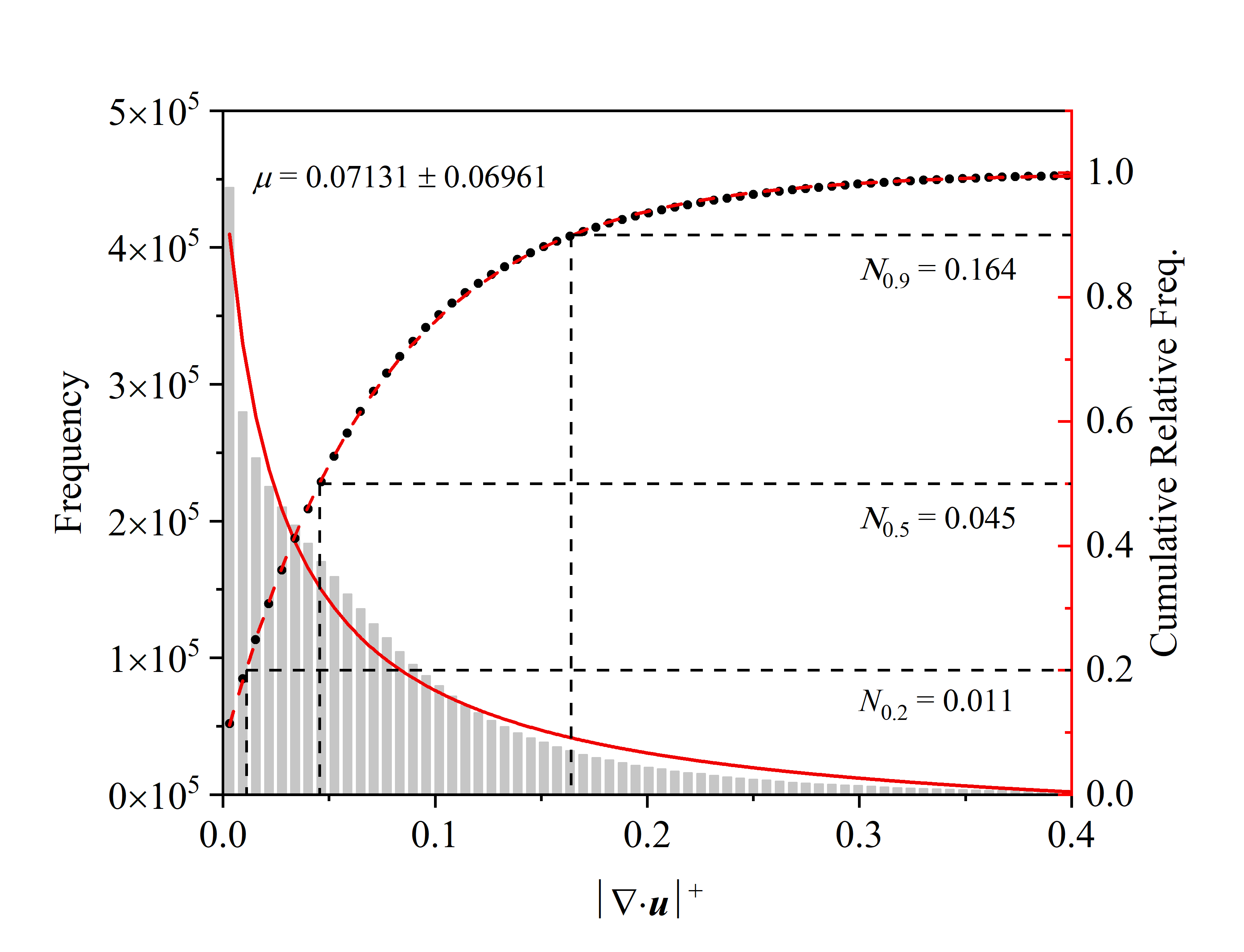}
  }
	\caption{Statistical distribution of divergence-free errors in the entire computation domain. (\textit{a}) Results of the spectrum adopting the nonuniform integral scale profile, and (\textit{b}) results of the spectrum adopting constant integral length scale.}
	\label{fig:cf:sta}
\end{figure}

Fig.\ref{fig:cf:sta} shows the statistical distribution of the divergence-free error of turbulence generated by the distributed integral scale and the constant one by a single realization. The bar chart shows the probability density distribution of the two choices. The red solid line is the approximate fitting line. First of all, it should be noted that no common distribution can accurately describe the probability distribution function of the two, and the relative frequency of both approach 0 rapidly, which reflects the effectiveness of the divergence-free correction. The black dot is the relative cumulative frequency of the divergence level of the probability distribution function, and the red dotted line is the fitting curve in an exponential form, and the figure also gives the specific divergence level at the cumulative frequency of 20\%, 50\%, and 90\%. The horizontal axis in the figure is the dimensionless result normalized by $u_\tau / \nu$. As exhibited in the figure, there are differences in the statistical distribution of divergence under the two integral scale choices: Although the results of the nonuniform integral scale in Fig.\ref{fig:cf:sta}(\textit{a}) have a larger frequency near 0, the frequency in the range of 0--0.5 is smaller than that of the constant integral scale in Fig.\ref{fig:cf:sta}(\textit{b}). But both the mean and the deviation are very close to each other. From the perspective of cumulative frequency, the divergence level of the varying scale at 20\% is slightly lower than the constant one, the level at 50\% is close, and the level at 90\% is slightly higher than the constant one. It can be seen that although there is a difference in the absolute divergence between the two, the effect is very small actually, and it is acceptable to ignore $Er_1$ when the distributed integral scale is adopted.

\begin{figure}
	\centering
  \fbox{
    \includegraphics[width=8cm]{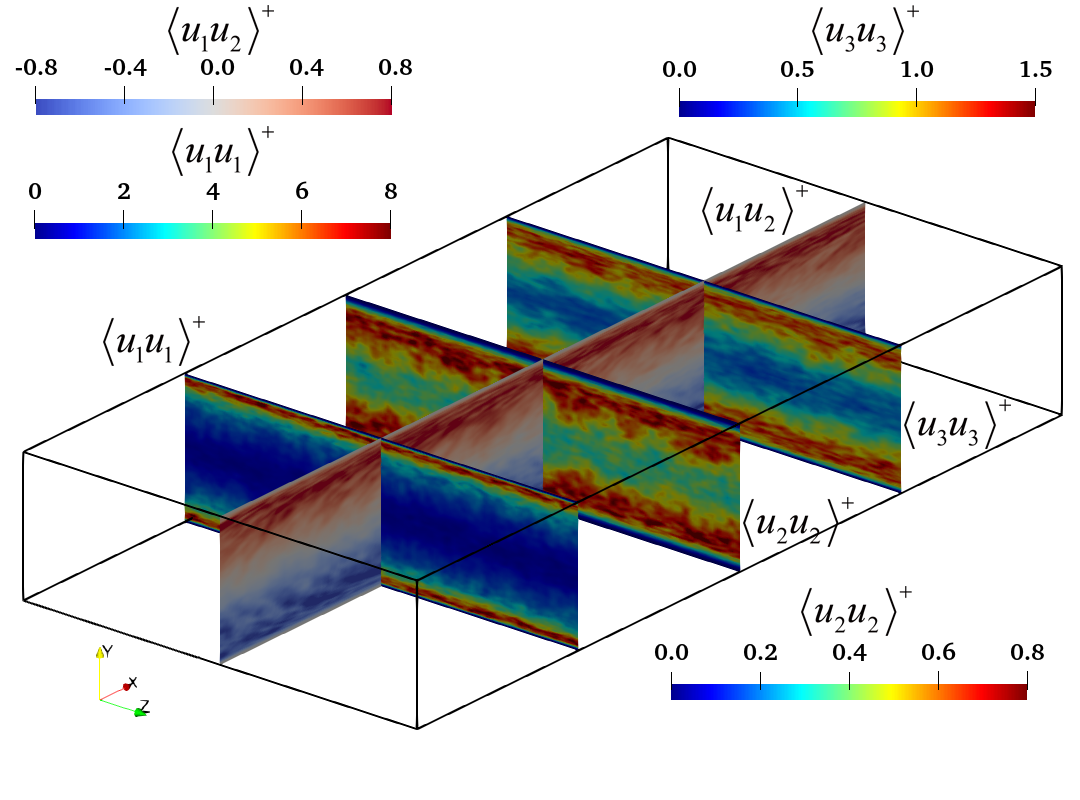}
    \includegraphics[width=8cm]{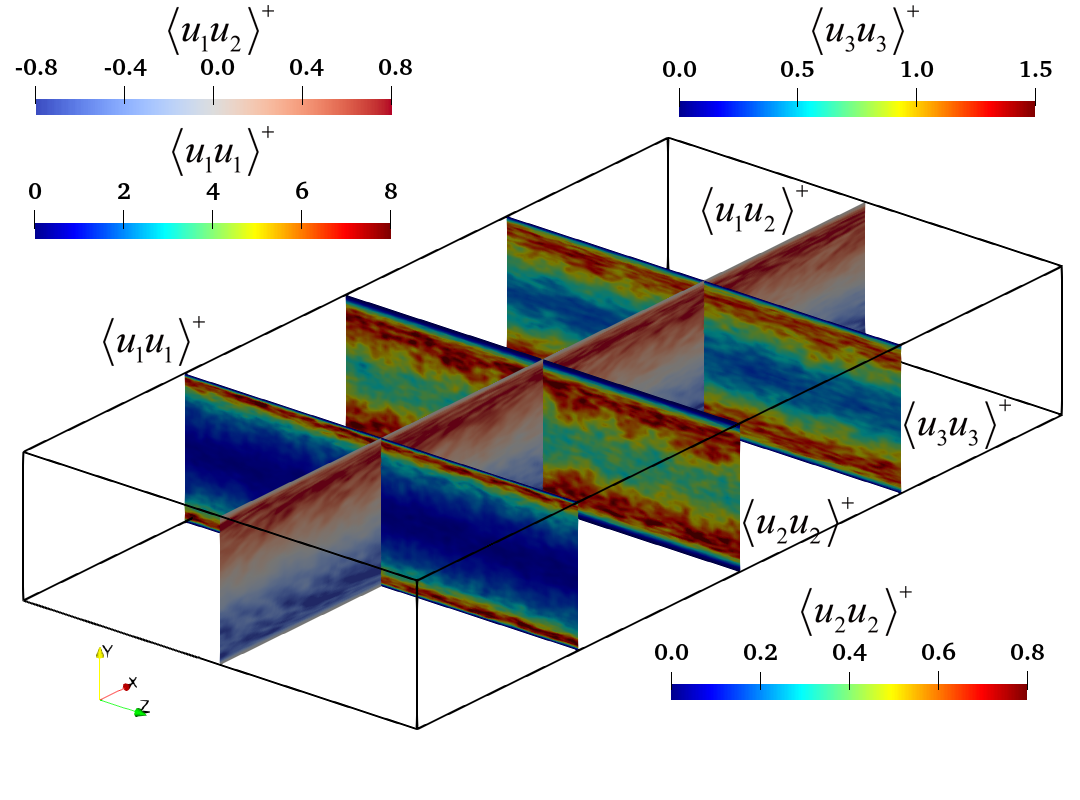}
  }
	\caption{Spatial distribution of ensemble-averaged Reynolds stress components in the channel flow (100 realizations). (\textit{a}) Results of the spectrum adopting the nonuniform integral scale profile, and (\textit{b}) results of the spectrum adopting constant integral length scale.}
	\label{fig:cfc}
\end{figure}

\begin{figure}
	\centering
  \fbox{
    \includegraphics[width=8cm]{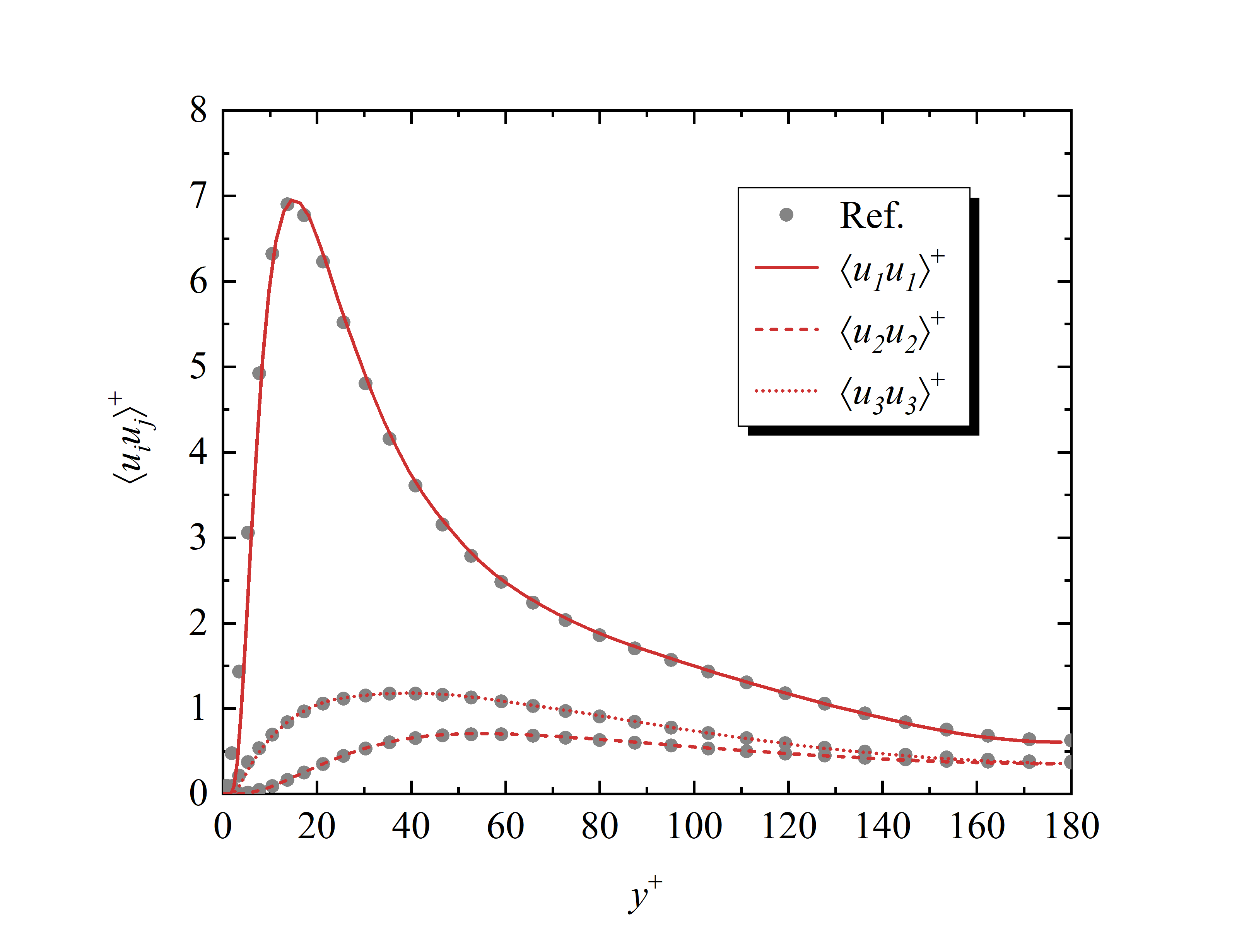}
    \includegraphics[width=8cm]{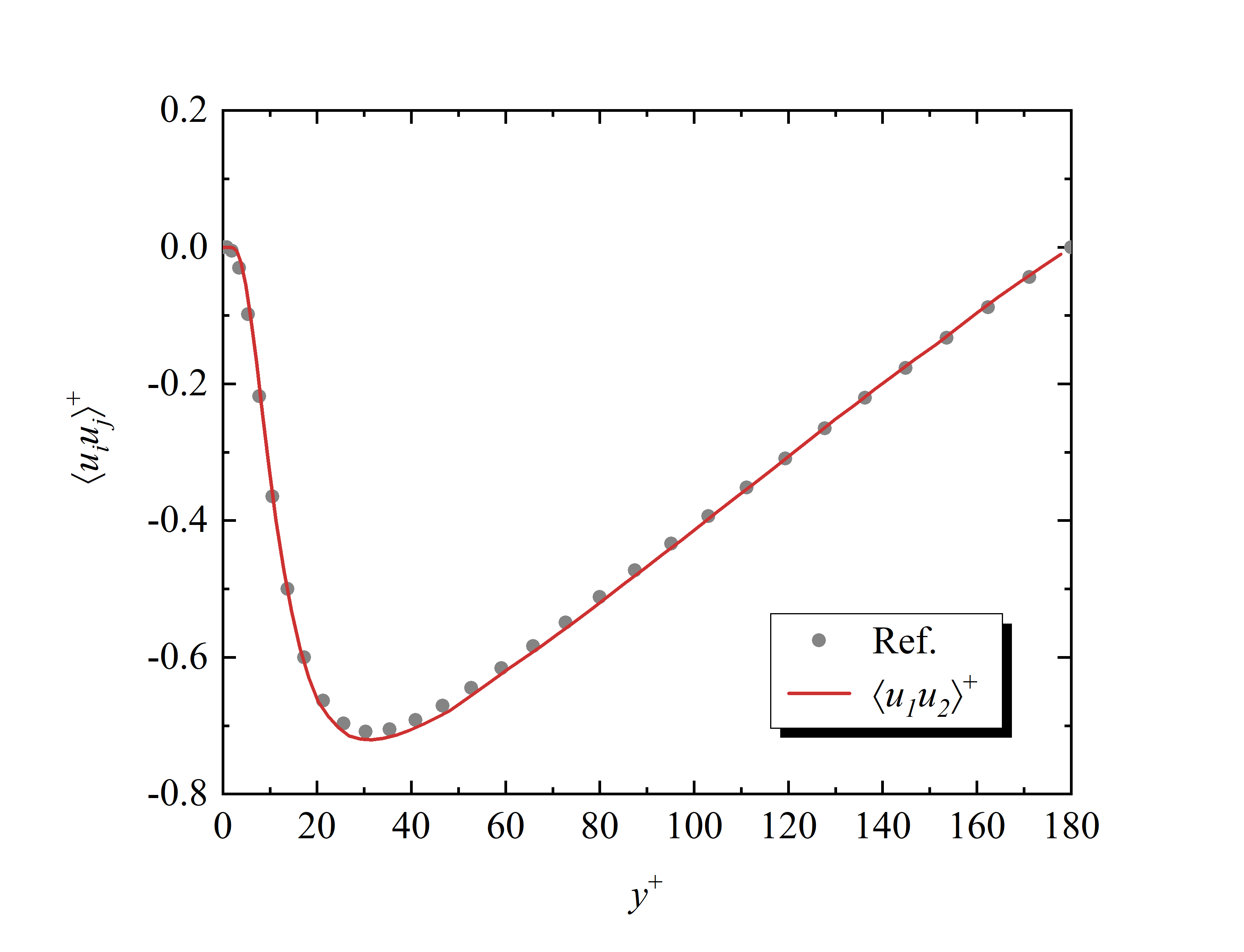}
  }
	\caption{The reproduction performance of stress profile by the proposed method (100 realizations, spatial averaging in x-z plane). (\textit{a}) normal stress, (\textit{b}) shear stress.}
	\label{fig:cf:stress}
\end{figure}

The generated correlation of two energy spectra using different integral scale models averaged using 100 samples are shown in Fig.\ref{fig:cfc}: the $z$ cross-section shows the shear stress contour while the other three planes show the normal stress one. The figure illustrates that the given correlation distribution is well produced by both scale models. It is worth noting that it is almost difficult to distinguish between Fig.\ref{fig:cfc}(\textit{a}) and Fig.\ref{fig:cfc}(\textit{b}), which indicates that the difference between the two is negligible and a constant integral scale model is accurate enough. The results of Fig.\ref{fig:cfc}(\textit{a}) are further spatially averaged in Fig.\ref{fig:cf:stress} to obtain normal stress profiles in Fig.\ref{fig:cf:stress}(\textit{a}) and the shear stress profile in Fig.\ref{fig:cf:stress}(\textit{b}), respectively, from which it can be seen that all profiles are in good agreement with benchmark data.

\begin{figure}
	\centering
  \fbox{
    \includegraphics[width=8cm]{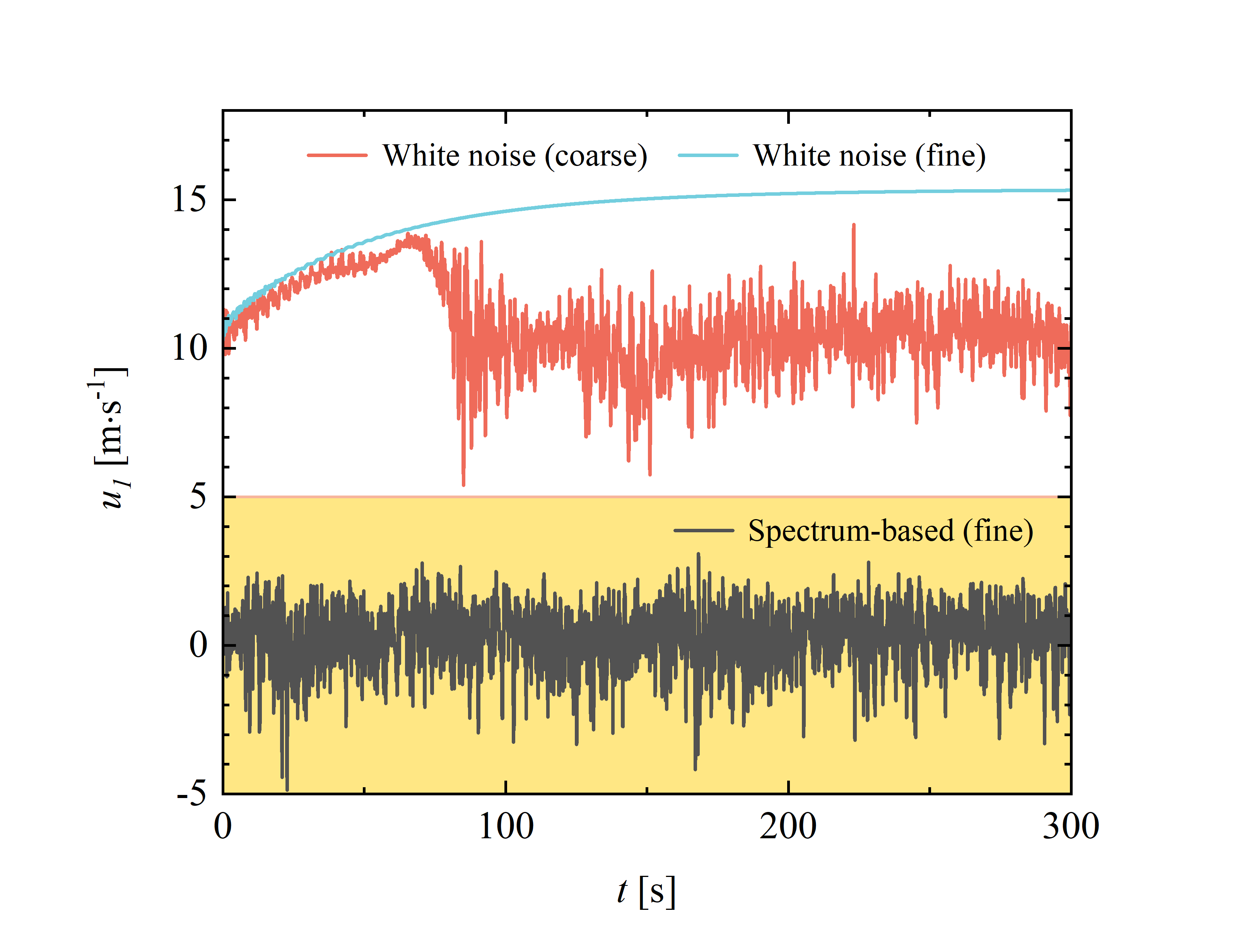}
    \includegraphics[width=8cm]{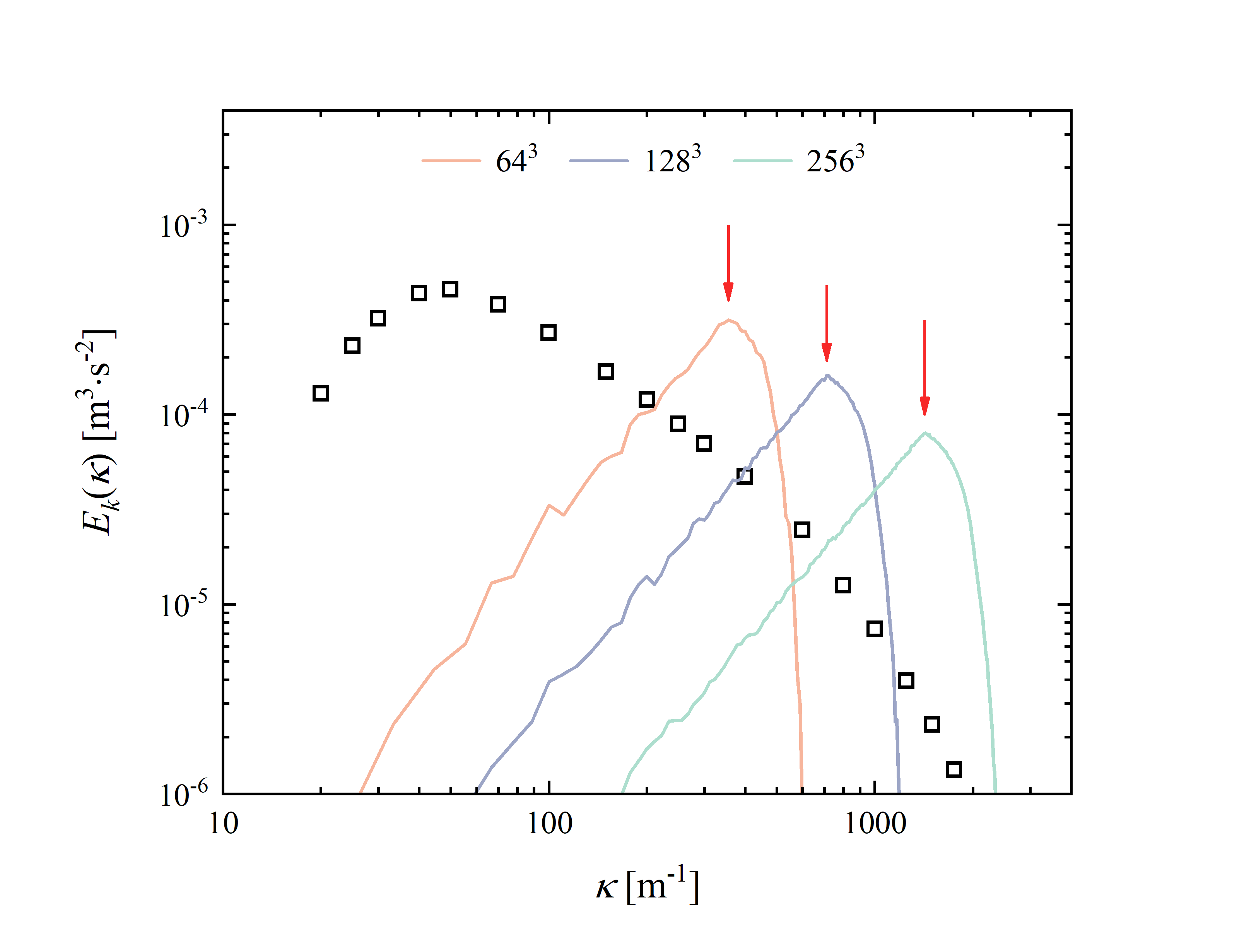}
  }
	\caption{The effectiveness of different types of synthetic turbulence methods when generating initial fields for LES computation. (\textit{a}) Comparison of turbulence decay between the proposed spectrum-based method and white noise generator in coarse and fine meshes, and (\textit{b}) energy spectra of white noise generator in different grid sizes with respect to the length scale.}
	\label{fig:cf:white}
\end{figure}

Finally, the new method is applied to the LES calculation of channel flow. To illustrate the turbulence decay problem, a comparison is made with the classical white noise generator. The calculation results for a coarser mesh are added to discuss the method performance under different mesh resolutions. Although Fig.\ref{fig:cf:white}(\textit{a}) only shows the instantaneous velocity at a probe near the core area, the instantaneous velocity and surface average Reynolds stress at other locations are also monitored and the behavior reflected by them is consistent. Fig.\ref{fig:cf:white}(\textit{a}) clearly shows that the turbulence initial field generated by the white noise method has a very obvious decay phenomenon in the CFD calculation, and this decaying exhibits a strong dependence on the gird size: the initial fluctuation under the coarse grid is strong, the decaying speed is slow, and the correct state can be regenerated by the grid itself. However, the initial fluctuation in the finer grid is weaker and the decaying process has taken place much faster. It is very difficult to generate turbulence again after initial turbulence becomes laminar totally (the figure only shows the calculation results for 300s, but there is still no fluctuation at all when the actual calculation time reaches 5000s!). In essence, this is caused by the lack of spatial correlation and the distortion of energy spectrum distribution in the white noise generated. Literature usually refers to this generator as the white noise method because the energy of the signal is the same at each frequency, but this conclusion only addresses the temporal characteristics of boundary turbulence. When analyzing the energy spectrum, we found that due to the spatial characteristics of turbulence, the energy spectrum of the white noise method has the behavior presented in Fig.\ref{fig:cf:white}(\textit{b}). The energy of large-scale turbulence was completely missing. Moreover, the smaller the grid size, the smaller the energy-containing scale, which means that for a grid with DNS resolution, the kinetic energy of the white noise method mostly falls into the dissipation scale. This feature leads to an unphysically larger dissipation of turbulence. The proposed spectrum-based method in Fig.\ref{fig:cf:white}(\textit{a}) is the inverter version one, barely any turbulence decaying phenomenon is observed, which shows the method's effectiveness as the initial field generator; The actual performance of the shifter version is very similar to that of the inverter one in Fig.\ref{fig:cf:white}(\textit{a}) (not shown here) indicating that it is also efficient as an initial field generation method even if anisotropy degeneration occurs.

\section{Conclusions}
\label{sec:conclusions}

In this paper, a high-efficient and low-divergence inhomogeneous and anisotropic turbulence generation method that can use arbitrary energy spectra is proposed. Verified by thorough test cases, the method can accurately restore the distribution of turbulence statistics and effectively correct divergence-free errors. The main conclusions are as follows.

(1) A high Reynolds number spectrum model is adopted in the proposed method. It is found in the verification that the non-uniform spectrum has almost no effect on the correlation function and divergence level of turbulence generated. Therefore, it is acceptable to use either the distributed integral scale or a constant one in practical applications.

(2) The shifter version of the method ensures a strict divergence-free feature in inhomogeneous and anisotropic turbulence, but issues of anisotropy degeneration and kinetic energy may occur. The latter can be effectively corrected by scaling functions. The existence of the former may require a larger recovery time or length. However, it is still effective to use the proposed method of this version as an initial field generator, because the degenerated anisotropy can quickly recover to the real state without unphysical turbulence decaying.

(3) To overcome degeneration issues and further increase the computation speed, the inverter version of the proposed method is constructed. Its effectiveness in anisotropy error correction has been well illustrated by several test cases. Although it leaves out the correction for inhomogeneity divergence-free error, this paper demonstrates through several verification cases that the increase in the divergence level in common applications is mainly due to anisotropy, and the error caused by non-uniformity is usually negligible. So for typical inhomogeneous and anisotropic turbulent flows in engineering, results obtained by the inverter version method are still both accurate and effective solenoidal.

(4) The proposed method uses the correlation reconstruction technique to solve the extension to the application of inhomogeneous and anisotropic turbulence, without any algorithm involving coordinate transformation or eigenvalue computations. Compared with spectrum-based methods proposed in previous studies, the new method has the advantages of high computational efficiency, less runtime memory cost, and easy implementation. At the same time, the required information is all stored locally, so it maintains the ability of multi-process parallelization in HPC, which is suitable for the practical computing requirements of large cost scale-resolving turbulence simulations like DNS and LES.

\section*{Code availability}

The source code to reproduce the results in this paper will be openly available on GitHub at \href{https://github.com/Fracturist/Synthetic-Turbulence-Generator}{Synthetic-Turbulence-Generator} upon publication.

\bibliographystyle{unsrtnat}
\bibliography{turbGenMe}

\end{document}